\documentclass[graybox]{svmult}

\usepackage{type1cm}        % activate if the above 3 fonts are
\usepackage{makeidx}         % allows index generation
\usepackage{graphicx}        % standard LaTeX graphics tool
\usepackage{multicol}        % used for the two-column index
\usepackage[bottom]{footmisc}% places footnotes at page bottom

\usepackage{newtxtext}       % 
\usepackage{newtxmath}       % selects Times Roman as basic font

\usepackage[utf8]{inputenc}
\usepackage{amsmath}
\usepackage{slashed}		%for Dirac slash  notation
\usepackage{bbm} 			%used to define identity matrix \idmat
\usepackage[breaklinks=true,colorlinks=true,linkcolor=blue,urlcolor=blue,citecolor=blue,bookmarks=true]{hyperref}
\def\!{\mskip-\thinmuskip}
\newcommand{\de}{\partial}										%  de: partial derivative
\renewcommand{\l}{\left}										%   l:	left
\renewcommand{\r}{\right}										%   r:	right
\newcommand{\mc}{\ensuremath{\mathcal}}        					%  mc:  calligrafico matematico
\newcommand{\h}{\widehat}                    	      			%   h:  hat
\newcommand{\ped}[1]{_{\textup{#1}}}							%pedices in math mode with up character
\newcommand{\apic}[1]{^{\textup{#1}}}							%apices in math mode with up character
\newcommand{\Dpi}{\mathcal{D}}									%Dpi: differential symbol for Path Integral
\newcommand{\parz}{Z}
\newcommand{\idmat}{\mathbb{I}}									%identity matrix variant
\newcommand{\group}[1]{\relax\ifmmode\mathsf{#1}\else\textsf{#1}\fi}	%element of a group in san serif
\newcommand{\tr}{{\text{tr}}}									% trace operator
\let\oldphi\phi \let\phi\varphi \let\varphi\oldphi		%invert \phi and \varphi
\newcommand{\mean}[1]{\langle #1 \rangle}						%statistical avarage
\newcommand{\oraw}[1]{\overrightarrow{#1}}
\newcommand{\olaw}[1]{\overleftarrow{#1}}
\newcommand{\mycorr}[1]{\langle\langle#1\rangle\rangle}
\hypersetup{pdfinfo={Title={Thermal chiral fermions with vorticity and magnetic field},Author={Matteo Buzzegoli}}}

\makeindex             % used for the subject index
\begin{document}

\title*{Thermodynamic equilibrium of massless fermions with vorticity, chirality and electromagnetic field}
\titlerunning{Thermal chiral fermions with vorticity and magnetic field}
\author{Matteo Buzzegoli}
\institute{Matteo Buzzegoli \at Universit\'a di Firenze and INFN Sezione di Firenze, Via G. Sansone 1, 
	I-50019 Sesto Fiorentino (Firenze), Italy, \email{buzzegoli@fi.infn.it}}
\maketitle

\abstract*{We present a study of the thermodynamics of the massless free Dirac
	field at equilibrium with axial charge, angular momentum and
	external electromagnetic field to assess the interplay between
	chirality, vorticity and electromagnetic field in relativistic fluids. 
	After discussing the general features of global thermodynamic
	equilibrium in quantum relativistic statistical mechanics, we 
	calculate the thermal expectation values. Axial imbalance and
	electromagnetic field are included non-perturbatively by using 
	the exact solutions of the Dirac equation, while a perturbative
	expansion is carried out in thermal vorticity. It is shown that 
	the chiral vortical effect and the axial vortical effect are not
	affected by a constant homogeneous electromagnetic field.}

\abstract{We present a study of the thermodynamics of the massless free Dirac
	field at equilibrium with axial charge, angular momentum and
	external electromagnetic field to assess the interplay between
	chirality, vorticity and electromagnetic field in relativistic fluids. 
	After discussing the general features of global thermodynamic
	equilibrium in quantum relativistic statistical mechanics, we 
	calculate the thermal expectation values. Axial imbalance and
	electromagnetic field are included non-perturbatively by using 
	the exact solutions of the Dirac equation, while a perturbative
	expansion is carried out in thermal vorticity. It is shown that 
	the chiral vortical effect and the axial vortical effect are not
	affected by a constant homogeneous electromagnetic field.}

%*********************************************************************************
\section{Introduction}
%*********************************************************************************
The collective macroscopic behavior of matter in the presence of quantum anomalies
and external fields is an increasingly important subject in several fields of physics.
Specifically, the experiments of relativistic heavy-ion collisions at RHIC and LHC
have posed new and interesting questions about the theoretical foundations of relativistic
collective phenomena. The experimental data of heavy ion collisions indicates the
creation of a deconfined quark gluon plasma in a strongly coupled regime at extreme
conditions of temperature, density, thermal vorticity~\cite{STAR:2017ckg} and magnetic
fields~\cite{Kharzeev:2015znc}. Moreover, it was argued~\cite{Kharzeev:2007jp,Fukushima:2008xe}
that the fluctuations of topological configurations of the QCD vacuum in the early stages
of a heavy ion collision generate a chiral imbalance, which is an imbalance between
the number of right- and left- handed quarks. Despite the usual relativistic
hydrodynamic has been very effective~\cite{Heinz:2013th} in reproducing the
experimental data for collective flow phenomena, it is now essential for the
interpretation of heavy ion collisions to address hydrodynamics in the contemporaneous
presence of chiral imbalance, thermal vorticity, and external electromagnetic fields.

The first crucial step towards understanding the hydrodynamics of matter subject to
external fields is to study its thermodynamic properties. It is the main purpose
of this contribution
to investigate the effects of an external electromagnetic field on the thermodynamics
of a chiral vorticous fluid. The effects of electromagnetic fields on (non-chiral
non-vorticous) relativistic quantum fluids were already studied in the past, see for
instance~\cite{HakimBook,Israel:1978up} and reference therein and~\cite{Canuto:1969cs}
for the special case of a constant magnetic field. More recently this topic has been
addressed in~\cite{Huang:2011dc} using the Zubarev's non-equilibrium statistical
operator, in~\cite{Kovtun:2016lfw} using the generating functional method and
in~\cite{Weickgenannt:2019dks,Gao:2019znl,Chen:2012ca} with Wigner function derived
from kinetic theory.

This contribution aims to highlight the modifications caused by chiral imbalance
and by thermal vorticity. The paper is organized in the following way.
In Sec.~\ref{sec:GlobEqEM}, we introduce the global thermal equilibrium of a chiral
system with the contemporaneous presence of an external electromagnetic field and a thermal
vorticity within Zubarev's non-equilibrium statistical operator formalism.
In Sec.~\ref{subsec:NoF} we give a brief overview of the main results for the case of
a chiral Dirac field in the absence of the electromagnetic field.
In Sec.~\ref{sec:DiracInB} we review the relativistic quantum theory of fermions
under the effect of an external magnetic field. Then, we obtain the exact form of the
chiral fermionic propagator with external constant magnetic field and we obtain the
exact thermal averages of the axial and electric currents.
In Sec.~\ref{sec:ThermoInConstantB} we examine the properties of a system at thermal
equilibrium with constant vorticity and electromagnetic field.
The last part of the paper is concerned with the consequences of an electromagnetic
field on the chiral vortical effect and the axial vortical effect.

%*********************************************************************************
\section*{Notation}
%*********************************************************************************
\label{sec:notation}
In this work we use the \emph{natural unit} system in which $\hslash=c=G=k_B=1$.
The \emph{Minkowski metric} is defined by the tensor $\eta_{\mu\nu}=$ diag$(1,-1,-1,-1)$;
for the Levi-Civita symbol we use the convention $\epsilon^{0123}=+1$.

Operators in Hilbert space will be denoted by a large upper hat, e.g. $\h{T}$ (with the exception 
of Dirac field operator that is denoted by $\Psi$). The stress-energy tensor used to define
Poincaré generators is always assumed to be symmetric with an associated vanishing spin tensor.

\newpage
%*********************************************************************************
\section{General global equilibrium with electromagnetic field}
%*********************************************************************************
\label{sec:GlobEqEM}
In this section we introduce the methods to study the thermodynamic equilibrium of a quantum relativistic system
in the presence of a chiral imbalance and of an external electromagnetic field. For
that purpose we review the Zubarev method of stationary non-equilibrium density
operator~\cite{Zubarev1979,vanWeert1982} (see also~\cite{BecaBetaF,Hayata:2015lga,Hongo:2016mqm,Becattini:2019dxo}
for recent developments) and we discuss the inclusion of a conserved axial current
and of an external electromagnetic field.
 
When we are dealing with a relativistic system we must consider local quantities
in order to address the appropriate covariant properties. To identify those quantities we use
a Arnowitt-Deser-Misner (ADM) decomposition of space-time~\cite{Zubarev1979,vanWeert1982,BecaBetaF,Hayata:2015lga}.
Choose then a foliation of space-time and suppose that the system in consideration thermalize faster than
the evolution of ``time'' $\tau$ in which we are interested. At each step of evolution $\D\tau$,
the system is at local thermal equilibrium and the macroscopic behavior of the system is described by
a stress-energy density $T_{\mu\nu}(x)$, an (electric) current density $j_\mu(x)$ and an axial current
${j\ped{A}}_\mu(x)$, all lying on a space-like hyper-surface $\Sigma(\tau)$. We can then describe
the thermal properties of the system with a density operator which lives on $\Sigma(\tau)$. As in the non-relativistic
case, the density operator at local equilibrium $\h\rho\ped{LE}$ is obtained as the operator
which maximizes the entropy $S=-\tr(\,\h\rho\log\h\rho\,)$. To reproduce the actual thermodynamics
on the hyper-surface we maximize the entropy with the constraints that the mean values of the stress-energy
tensor and of the currents on $\Sigma(\tau)$ corresponds to the values of the densities
$T_{\mu\nu}(x)$, $j_\mu(x)$ and ${j\ped{A}}_\mu(x)$~\cite{vanWeert1982}. To obtain these densities
we project the stress–energy tensor and the current mean values onto $n$, i.e. the normalized
four-vector perpendicular to~$\Sigma$:
\begin{equation*}
%\label{eq:ConsRel}
\begin{split}
n_\mu(x)\,\tr\left[\,\h{\rho}\,\,\h{T}^{\mu\nu}(x)\right]=&\,n_\mu(x)\,\mean{\,\h{T}^{\mu\nu}(x)}\equiv n_\mu(x)\,T^{\mu\nu}(x),\\
n_\mu(x)\,\tr\left[\,\h{\rho}\,\,\h{j}^{\,\mu}(x)\right]=&\,n_\mu(x)\,\mean{\h{j}^{\,\mu}(x)}\equiv n_\mu(x)\,j^\mu(x),\\
%n_\mu(x)\,\tr\left[\,\h{\rho}\,\,\h{j}\ped{A}^{\,\mu}(x)\right]=&\,n_\mu(x)\,\mean{\h{j}\ped{A}^{\,\mu}(x)}\equiv n_\mu(x)\,j\ped{A}^\mu(x).
\end{split}
\end{equation*}
and similarly for the axial current.
We could also impose a constraint on the angular momentum density, but since we are choosing the Belinfante
operator as the stress–energy tensor, it turns out that this additional requirement is automatically
taken into account~\cite{Becattini:2018duy}.

The maximum solution $\h{\rho}\ped{LE}$ gives the \emph{Local Equilibrium Density Operator} (LEDO)~\cite{BecaBetaF,Hayata:2015lga}:
\begin{equation}
\label{eq:rhoLTEF}
\h{\rho}\ped{LTE} = \frac{1}{\parz} \exp \left[ -\int_{\Sigma} \D \Sigma_\mu \left( \h T^{\mu\nu}(x) \beta_\nu(x) - \zeta(x)\, \h j^{\,\mu}(x)
- \zeta\ped{A}(x)\, \h j\ped{A}^{\,\mu}(x)\right) \right],
\end{equation}
where $\beta^\mu$ is the four-temperature vector such that $T=1/\sqrt{\beta^2}$  is the proper
comoving temperature, $\zeta$ and $\zeta\ped{A}$ are the ratio of comoving chemical potentials and
the temperature (e.g. $\zeta=\mu/T$) and $\parz$ is the partition function.
In the presence of an external electromagnetic field, we indicate with $A^\mu(x)$ the non-dynamical
gauge field and with $F^{\mu\nu}=\de^\mu A^\nu-\de^\nu A^\mu$ the electromagnetic strength tensor.
Therefore, the operator relations stemming for conservation equations are
\begin{equation}
\label{eq:OperRelF}
\de_\mu \h j^{\,\mu}=0,\quad \de_\mu\h T^{\mu\nu}=\h j_\lambda F^{\nu\lambda},\quad\de_\mu \h j^{\,\mu}\ped{A}=0.
\end{equation}
Furthermore, the four-momentum operator $\h P$ and the conserved charges $\h Q_i$ are obtained by
\begin{equation*}
\h P^\mu=\int_\Sigma \D\Sigma_\lambda\, \h T^{\lambda\mu},\qquad
\h Q_i=\int_\Sigma \D\Sigma_\lambda\, \h j^{\,\lambda}_i,
\end{equation*}
while the angular momentum is
\begin{equation}
\label{Lorentzgen}
\h J^{\mu\nu}=\int_\Sigma \D\Sigma_\lambda \left(x^\mu\h T^{\lambda\nu}-x^\nu\h T^{\lambda\mu}\right).
\end{equation}
Notice that, as we will discuss in details in Sec.~\ref{sec:DiracInB}, the four-momentum $\h P$
and the angular momentum $\h J$ in the presence of external electromagnetic field are neither
conserved nor the generators of translations and Lorentz transformations.

In the case of Dirac fermions interacting with an external gauge field, the explicit
form of the operators above is
\begin{equation}
\label{eq:DiracOperator}
\begin{split}
\h j^{\,\mu} =&\, q\bar{\Psi} \gamma^\mu \Psi, \quad \h j^{\,\mu}\ped{A} =\, \bar{\Psi} \gamma^\mu\gamma^5 \Psi,\\
\h T^{\mu\nu}= &\frac{\I}{4}\left[\bar{\Psi}\gamma^\mu\oraw{\de}^\nu\Psi-\bar{\Psi}\gamma^\mu\olaw{\de}^\nu\Psi
	+\bar{\Psi}\gamma^\nu\oraw{\de}^\mu\Psi -\bar{\Psi}\gamma^\nu\olaw{\de}^\mu\Psi\right]
	- \frac{1}{2}\left(\h j^{\,\mu} A^\nu+\h j^\nu A^\mu\right),
\end{split}
\end{equation}
and the stress-energy tensor and the electric current indeed satisfy the relations~(\ref{eq:OperRelF}).
Regarding the axial current $\h{j}\ped{A}$, we also have to take into account the chiral anomaly.
The chiral anomaly affects the axial current divergence as follows
\begin{equation*}
\de_\mu \h{j}^{\,\mu}\ped{A}=-\frac{1}{8}\epsilon^{\mu\nu\rho\lambda}\frac{q^2}{2\pi^2}F_{\mu\nu} F_{\rho\lambda}
	=-\frac{q^2}{2\pi^2}(E\cdot B),
\end{equation*}
where $q$ is the electric charge of the fermion, and $E$ and $B$ are comoving electric and
magnetic field, defined by
\begin{equation*}
F^{\mu\nu}=E^\mu u^\nu -E^\nu u^\mu - \epsilon^{\mu\nu\rho\sigma} B_\rho u_\sigma,
\end{equation*}
with $u$ the fluid velocity. Even when the product $E\cdot B$ is
non-vanishing and consequently the axial current is not conserved, we can still define
a new conserved ``axial'' current by means of the Chern-Simons current $K$,
whose divergence gives the chiral anomaly:
\begin{equation*}
K^\mu=\epsilon^{\mu\nu\rho\sigma}A_\nu F_{\rho\sigma},\quad
\frac{q^2}{8\pi^2}\de_\mu K^\mu=\frac{q^2}{2\pi^2}(E\cdot B).
\end{equation*}
The new conserved axial current $\h{j}\ped{CS}$ is then defined as
\begin{equation*}
\h{j}^{\,\mu}\ped{CS}=\h{j}^{\,\mu}\ped{A}+\frac{q^2}{8\pi^2}K^\mu,\quad
\de_\mu \h{j}^{\,\mu}\ped{CS}=0,
\end{equation*}
and the axial chemical potential $\mu\ped{A}$ is to be associated to this current.
Since the additional current $K$ depends only on external fields, it is not
a quantum operator and it does not contribute to thermal averages. Therefore, all the
results discussed in the absence of chiral anomaly will also be valid for the case
of equilibrium with conserved Chern-Simons current. Because there is no difference
in the results, we will continue to denote the current associated to $\mu\ped{A}$
inside the statistical operator with $\h{j}\ped{A}$ even when the chiral anomaly
is non-vanishing.

Let us now move to describe the system at global thermal equilibrium.
The global equilibrium is reached when the statistical operator~(\ref{eq:rhoLTEF})
is time independent. This occurs when the integrand inside Eq.~(\ref{eq:rhoLTEF})
is divergenceless~\cite{BecCov}. Then, it is easily proven using relations~(\ref{eq:OperRelF})
that global thermal equilibrium is realized when the following relations are satisfied:
\begin{equation}
\label{eq_GlobEqCond}
\de_{\mu}\beta_{\nu}(x)+\de_{\nu}\beta_{\mu}(x)=0,\qquad \de^\mu \zeta(x)=F^{\nu\mu}\beta_\nu(x)
,\qquad \de^\mu \zeta\ped{A}(x)=0.
\end{equation}
The inverse four-temperature and the axial chemical potential solves the previous
conditions if they are given by~\cite{BecaBetaF}:
\begin{equation*}
\beta_\mu(x)=b_\mu+\varpi_{\mu\rho}x^\rho,\quad \zeta\ped{A}=\text{constant},
\end{equation*}
where $b$ is a constant time-like four-vector and $\varpi$ is a constant anti-symmetric tensor.
We refer to $\varpi$ as the \emph{thermal vorticity} because it is the anti-symmetric derivative
of inverse four-temperature:
\begin{equation*}
\varpi_{\mu\nu}=-\frac{1}{2}\left(\de_\mu\beta_\nu-\de_\nu\beta_\mu \right)
\end{equation*}
and because it contains information about the fluid's acceleration and rotation.
Indeed, if the $\beta$ four-vector is a time-like vector, then we
can choose the $\beta$-frame as hydrodynamic frame~\cite{BecCov,Kovtun:2016lfw}. 
The unitary four-vector fluid velocity $u$ is therefore identified with the direction of $\beta$:
\begin{equation*}
u^\mu(x)=\frac{\beta^\mu(x)}{\sqrt{\beta_\rho(x)\beta^\rho(x)}}.
\end{equation*}
As long as we are considering physical observables in a region where the coordinate $x$ are
such that $\beta(x)$ is a time-like vector, this definition provides a proper
choice for the fluid velocity.
We can decompose the thermal vorticity into two space-like vector fields, each having three
independent components, by projecting along the time-like fluid velocity $u$:
\begin{equation*}
\varpi^{\mu\nu}=\epsilon^{\mu\nu\rho\sigma}w_\rho u_\sigma+\alpha^\mu u^\nu-\alpha^\nu u^\mu.
\end{equation*}
The four-vectors $\alpha$ and $w$ are explicitly written inverting the previous relation: 
\begin{equation*}
%\label{eq:accvort}
\alpha^\mu(x)\equiv\varpi^{\mu\nu}u_\nu,\quad w^\mu(x)
	\equiv-\frac{1}{2}\epsilon^{\mu\nu\rho\sigma}\varpi_{\nu\rho}u_\sigma.
\end{equation*}
The vectors $\alpha$ and $w$ depend on the coordinates, are space-like, and are orthogonal to $u$.
All the quantity $u,\,\varpi,\,\alpha,\,w$ are dimensionless. 
From their definitions, we can easily show that $\alpha$ and $w$ are given by
\begin{equation*}
\alpha^\mu=\sqrt{\beta^2}\,a^\mu,\quad w^\mu=\sqrt{\beta^2}\,\omega^\mu,
\end{equation*}
where $a$ and $\omega$ are the local acceleration and rotation of the fluid,
which are given by
\begin{equation*}
a^\mu=u_\nu\de^\nu u^\mu,\quad\omega^\mu\equiv\frac{1}{2}\epsilon^{\mu\nu\rho\sigma}u_\sigma \de_\nu u_\rho.
\end{equation*}
Furthermore, it will proves useful to define the projector into the orthogonal space of fluid velocity:
\begin{equation*}
\Delta_{\mu\nu}\equiv g_{\mu\nu}-u_\mu u_\nu,
\end{equation*}
and the four vector $\gamma$ orthogonal to the other ones: $u,\alpha,w$
\begin{equation*}
%\label{eq:transversedir}
\gamma^\mu=\epsilon^{\mu\nu\rho\sigma}w_\nu\alpha_\rho u_\sigma=(\alpha\cdot \varpi)_\lambda \Delta^{\lambda\mu}.
\end{equation*}
The $\varpi$ decomposition above defines a tetrad $\{u,\alpha,w,\gamma\}$ which can be used as a basis for four-vectors.
It must be noticed, however, that the tetrad is neither unitary nor orthonormal, indeed in general we
have $\alpha\cdot w\neq 0$.

Returning to the global equilibrium conditions \eqref{eq_GlobEqCond}, notice that in the absence of
electromagnetic field also $\zeta$ must be a constant. In that case the global equilibrium
statistical operator takes the following form~\cite{BecGro,Buzzegoli:2018wpy}:
\begin{equation}
\label{eq:StatGEnoF}
\h\rho=\frac{1}{\parz}\exp\left\{-b\cdot\h{P}+\frac{1}{2}\varpi:\h{ J}+\zeta\h{Q}+\zeta\ped{A}\h{Q}\ped{A}\right\}.
\end{equation}
The thermodynamics of Dirac fermions which follows from this operator is quickly
reviewed in Section~\ref{subsec:NoF}

For the case of a non-vanishing electromagnetic field instead, we need to solve the equation:
\begin{equation}
\label{eq_ZetaGeneral}
\de^\mu \zeta(x)= F^{\sigma\mu}\beta_\sigma.
\end{equation}
To find the solution, we first derive it respect to $\de^\nu$:
\begin{equation}
\label{eq_derZetaGeneral}
\de^\nu\de^\mu \zeta=\de^\nu( F^{\sigma\mu}\beta_\sigma).
\end{equation}
Since we can exchange the order of the derivatives $\de^\nu\de^\mu$ on the l.h.s. of \eqref{eq_derZetaGeneral},
it follows that the anti-symmetrization respect to indices $\mu$ and $\nu$ of \eqref{eq_derZetaGeneral} must be vanishing:
\begin{equation*}
\begin{split}
\de^\nu \de^\mu \zeta -\de^\mu \de^\nu \zeta& =0=\left[\de^\nu( F^{\sigma\mu}\beta_\sigma)
	-\de^\mu( F^{\sigma\nu}\beta_\sigma)\right]\\
&=\left[\beta_\sigma(\de^\nu F^{\sigma\mu}-\de^\mu F^{\sigma\nu})
	+(\de^\nu\beta_\sigma) F^{\sigma\mu}+(\de^\mu\beta_\sigma) F^{\nu\sigma}\right].
\end{split}
\end{equation*}
Using the first Bianchi identity $\de^\nu F^{\sigma\mu}+\de^\mu F^{\nu\sigma}+\de^\sigma F^{\mu\nu}=0$, we obtain
\begin{equation*}
\beta_\sigma \de^\sigma F^{\mu\nu}+(\de^\nu\beta_\sigma) F^{\mu\sigma}+(\de^\mu\beta_\sigma) F^{\sigma\nu}=0.
\end{equation*}
We may recognize the Lie derivative of $F$ along $\beta$ in the previous equation. This constitutes a first
condition for global equilibrium, the system can reach global equilibrium only when
\begin{equation}
\label{eq_EqConstraintFmunu}
\mc{L}_\beta (F)=0,\quad\leftrightarrow\quad
\beta_\sigma(x)\de^\sigma F^{\mu\nu}(x)=\varpi^\mu_{\hphantom{\mu}\sigma} F^{\sigma\nu}(x)
	-\varpi^\nu_{\hphantom{\nu}\sigma} F^{\sigma\mu}(x),
\end{equation}
that is to say when the electromagnetic field follows the field lines of inverse-four temperature.

To actually solve Eq.~(\ref{eq_ZetaGeneral}) we translate the global equilibrium
condition of the strength tensor~(\ref{eq_EqConstraintFmunu}) to the four-vector
potential $A^\mu$. We see that the constraint (\ref{eq_EqConstraintFmunu}) is
satisfied if $A$ solves
\begin{equation}
\label{eq_EqConstraintAmu}
\beta_\sigma(x)\de^\sigma A^\mu(x)=\varpi^\mu_{\hphantom{\mu}\sigma}A^\sigma(x)+\de^\mu\Phi(x),
\end{equation}
where $\Phi$ is a smooth function of $x$. In ref.~\cite{Hongo:2016mqm} was also stated
that a gauge potential with vanishing Lie derivative along $\beta$ gives a stationary
statistical operator, which is condition~\eqref{eq_EqConstraintAmu}. It is important
to stress that after a gauge transformation the condition~(\ref{eq_EqConstraintAmu})
still holds true for the new gauge potential because the function $\Phi$ is also
affected by the gauge transformation. Indeed, let $A^\mu$ satisfies Eq.~(\ref{eq_EqConstraintAmu});
after the gauge transformation $A^{\prime\mu}=A^\mu+\de^\mu\Lambda$ we find:
\begin{equation*}
\begin{split}
\beta_\sigma\de^\sigma A^{\prime\mu}=&\beta_\sigma \de^\sigma A^\mu+\beta_\sigma\de^\sigma \de^\mu\Lambda
=\omega^\mu_{\,\,\sigma}A^\sigma+\de^\mu\Phi+\de^\mu(\beta_\sigma\de^\sigma\Lambda)-(\de^\mu\beta_\sigma)\de^\sigma\Lambda\\
=&\varpi^\mu_{\hphantom{\mu}\sigma}(A^\sigma+\de^\sigma\Lambda)+\de^\mu(\Phi+\beta_\sigma\de^\sigma\Lambda)
=\varpi^\mu_{\hphantom{\mu}\sigma}A^{\prime\sigma}+\de^\mu \Phi',
\end{split}
\end{equation*}
which is exactly condition~(\ref{eq_EqConstraintAmu}) for $A^{\prime\mu}$ and for $\Phi'$,
that is $\Phi$ shifted by the transport of $\Lambda$ along $\beta$.

We can now write Eq.~(\ref{eq_ZetaGeneral}) by taking advantage of Eq.~(\ref{eq_EqConstraintAmu}):
\begin{equation*}
\begin{split}
\de^\mu \zeta &= F^{\sigma\mu}\beta_\sigma=\beta_\sigma (\de^\sigma A^\mu-\de^\mu A^\sigma)
=\beta_\sigma \de^\sigma A^\mu -\de^\mu (\beta_\sigma A^\sigma)+(\de^\mu \beta_\sigma)A^\sigma\\
&= \varpi^\mu_{\,\sigma}A^\sigma +\de^\mu\Phi- \varpi^\mu_{\,\sigma}A^\sigma-\de^\mu(\beta_\sigma A^\sigma).
\end{split}
\end{equation*}
We can then collect all the derivatives together into the equation
\begin{equation*}
\de^\mu\left(\zeta-\Phi+\beta_\sigma A^\sigma\right)=0,
\end{equation*}
from which we immediately get the solution:
\begin{equation}
\label{eq_ZetaEmEq}
\zeta(x)=\zeta_0-\beta_\sigma(x)A^\sigma(x)+\Phi(x),
\end{equation}
where $\zeta_0$ is a constant. The parameter $\Phi$ is analogous to the parameter which grants gauge invariance to
chemical potential in~\cite{Jensen:2013kka}. Even though Eq.~(\ref{eq_ZetaEmEq}) is given in terms of the gauge potential it
is still gauge invariant. Indeed, we have shown that with a gauge transformation, $A^\mu$ and $\Phi$ transform as
\begin{equation*}
A^{\prime\mu}=A^\mu+\de^\mu\Lambda,\quad \Phi'=\Phi+\beta_\sigma\de^\sigma\Lambda,
\end{equation*}
therefore, the chemical potential $\zeta$ is overall unaffected by gauge transformations:
\begin{equation*}
\zeta(x)'=\zeta_0-\beta_\sigma(x)A^{\prime\sigma}+\Phi'
=\zeta_0-\beta_\sigma(x)A^{\sigma}+\Phi-\beta_\sigma\de^\sigma\Lambda+\beta_\sigma\de^\sigma\Lambda
=\zeta(x).
\end{equation*}
The global equilibrium statistical operator is then obtained from the local one in Eq.~(\ref{eq:rhoLTEF})
by replacing the global equilibrium form of the thermodynamic fields $\beta,\zeta,\zeta\ped{A}$:
\begin{equation}
\label{eq:GEDO_EM}
\begin{split}
\h{\rho} =& \frac{1}{Z} \exp \left\{ -\int_{\Sigma} \D \Sigma_\mu \left[ \left(\h T^{\mu\nu}(x) + \h j^{\,\mu}(x) A^\nu(x)\right)\beta_\nu(x)\right.\right.\\
&\left.\left.- \left(\zeta_0+\Phi(x)\right) \h j^{\,\mu}(x) - \zeta\ped{A}\, \h j\ped{A}^{\,\mu}(x)\right] \right\}.
\end{split}
\end{equation}
This operator, on par with (\ref{eq:StatGEnoF}), is given in terms of global conserved quantities.
The difference is that for the general form of the external magnetic field  satisfying the constraint
(\ref{eq_EqConstraintFmunu}), the integration over the hyper-surface $\Sigma$ does not give easily
recognizable quantities like the four-momentum and the angular momenta in Eq.~(\ref{eq:StatGEnoF}).
However, the identification of global conserved operators can be carried out in the special case
of constant homogeneous electromagnetic field and it is discussed in the following sections.

%*********************************************************************************
\subsection{Vanishing electromagnetic field}
%*********************************************************************************
\label{subsec:NoF}
Before proceeding with the effects of electromagnetic fields, we briefly review the
thermodynamics properties of a relativistic system in the presence of thermal vorticity
but without an external electromagnetic field.
Regarding thermal equilibrium in the presence of rotation, exact solutions for
the free scalar and Dirac fields are discussed in~\cite{Ambrus:2014uqa,Ambrus:2019cvr,Becattini:2020qol}.
Instead the effects of acceleration has been recently investigated using the Zubarev method
in Ref.~\cite{Becattini:2020qol,Becattini:2017ljh,Becattini:2019poj,Prokhorov:2019sss,Prokhorov:2019cik,Prokhorov:2019yft}.
Here we want to report the constitutive equations at second order on thermal vorticity
discussed in~\cite{BecGro,Buzzegoli:2017cqy} and in \cite{Buzzegoli:2018wpy} including an axial current
(see also \cite{Hongo:2019rbd} for first order in thermal vorticity and magnetic field).
Using linear response theory on thermal vorticity, the thermal expectation value of a local
operator $\h O(x)$ evaluated with statistical operator~(\ref{eq_EqConstraintFmunu}) can
be written as~\cite{BecGro,Buzzegoli:2018wpy}:
\begin{equation}
\label{eq:VortExp}
\begin{split}
\mean {\h O(x)}=&\mean{\h O(0)}_{\beta(x)}-\alpha_\rho \mycorr{\,\h K^\rho \h O\,}-w_\rho \mycorr{\,\h J^\rho \h O\,}
	+\frac{\alpha_\rho\alpha_\sigma}{2}\mycorr{\,\h K^\rho \h K^\sigma \h O\,}\\
&+\frac{w_\rho w_\sigma}{2} \mycorr{\,\h J^\rho \h J^\sigma \h O\,}
	+\frac{\alpha_\rho w_\sigma}{2}\mycorr{\,\{\h K^\rho,\h J^\sigma\}\h O\,}+\mathcal{O}(\varpi^3).
\end{split}
\end{equation}
In the previous expression we indicated with double angular bracket the correlator
\begin{equation*}
\begin{split}
\mycorr{\h K^{\rho_1}\cdots \h K^{\rho_n} \h J^{\sigma_1}\cdots \h J^{\sigma_m}& \h O} \equiv
	\int_0^{|\beta|} \frac{\D\tau_1\cdots\D\tau_{n+m}}{|\beta|^{n+m}}\times\\
&\times\mean{{\rm T}_\tau\left(\h K^{\rho_1}_{-\I \tau_1 u}\cdots\h K^{\rho_n}_{-\I \tau_n u}
	\h J^{\sigma_1}_{-\I \tau_{n+1} u} \cdots\h J^{\sigma_m}_{-\I \tau_{n+m} u} \h O(0)\right)}_{\beta(x),c},
\end{split}
\end{equation*}
where $\h{J}$ and $\h{K}$ are the comoving rotation and boost generators, identified by
\begin{equation*}
\h{K}^\mu=u_\lambda\h{J}^{\lambda\mu},\quad
\h{J}^\mu=\frac{1}{2}\epsilon^{\alpha\beta\gamma\mu}u_\alpha\h{J}_{\beta\gamma},
\end{equation*}
and the averages $\mean{\cdots}_{\beta(x)}$ are evaluated at a fixed point $x$ with the
homogeneous statistical operator
\begin{equation*}
\h\rho_0=\frac{1}{\parz_0}\exp\l\{-\beta(x)\cdot\h{P}+\zeta(x)\h{Q}+\zeta\ped{A}(x)\h{Q}\ped{A}\r\}.
\end{equation*}
The subscript $-\I\tau u$ indicates an imaginary translation along $u$ as follows
\begin{equation*}
J^{\mu}_{-\I \tau u}\equiv \E^{-\I \tau u\cdot \h P} \h J^\mu \E^{\I \tau u\cdot \h P}.
\end{equation*}
Constitutive equations at second order on thermal vorticity of the stress-energy tensor,
the electric current and the axial current can be obtained using the expansion in Eq.~(\ref{eq:VortExp}).
We obtain~\cite{BecGro,Buzzegoli:2017cqy,Buzzegoli:2018wpy}
\begin{equation}
\begin{split}\label{setdecomp}
\mean{\h T^{\mu\nu}}=&\,\mathbb{A}\,\epsilon^{\mu\nu\kappa\lambda}\alpha_\kappa u_\lambda+\mathbb{W}_1 w^\mu u^\nu +\mathbb{W}_2 w^\nu u^\mu\\
&+(\rho-\alpha^2 U_\alpha -w^2 U_w)u^\mu u^\nu -(p-\alpha^2D_\alpha-w^2D_w)\Delta^{\mu\nu}\\
&+A\,\alpha^\mu\alpha^\nu+Ww^\mu w^\nu+G_1 u^\mu\gamma^\nu+G_2 u^\nu\gamma^\mu+\mathcal{O}(\varpi^3),
\end{split}
\end{equation}
\begin{equation}\label{vcurrdecomp}
\mean{\h j\ped{V}^\mu}=n\ped{V}\,u^\mu+\left(\alpha^2 N\apic{V}_\alpha+w^2 N\apic{V}_\omega\right)u^\mu+W\apic{V}w^\mu+G\apic{V}\gamma^\mu
+\mathcal{O}(\varpi^3),
\end{equation}
\begin{equation}\label{acurrdecomp}
\mean{\h j\ped{A}^\mu}=n\ped{A}\,u^\mu+\left(\alpha^2 N_\alpha\apic{A}+w^2 N_\omega\apic{A}\right)u^\mu+W\apic{A}w^\mu+G\apic{A}\gamma^\mu
+\mathcal{O}(\varpi^3).
\end{equation}
Not all of these coefficients are independent, indeed conservation equations (\ref{eq:OperRelF}) impose
the following relations~\cite{Buzzegoli:2017cqy} (this is explained in details in Sec.~\ref{subsec:Feffect}):
\begin{equation*}
%\label{eq:setrel}
\begin{split}
U_\alpha&=-|\beta|\frac{\partial}{\partial|\beta|}\big(D_\alpha+A\big)-\big(D_\alpha+A\big),\\
U_w&=-|\beta|\frac{\partial}{\partial|\beta|}\big(D_w+W\big)-D_w+2A-3W,\\
G_1+G_2&=2\big(D_\alpha+D_w\big)+A+|\beta|\frac{\partial}{\partial|\beta|}W+3W,
\end{split}
\end{equation*}
instead, for the first-order coefficients, conservation equations require that
\begin{equation*}
%\label{setchiralrel}
-2\mathbb{A}=|\beta|\frac{\de \mathbb{W}_1}{\de|\beta|}+3\mathbb{W}_1+\mathbb{W}_2.
\end{equation*}
For electric and axial current, we find that only the following equations must be fulfilled:
\begin{equation}\label{AVERel}
|\beta|\frac{\de W\apic{V}}{\de |\beta|}+3W\apic{V}=0,\quad
|\beta|\frac{\de W\apic{A}}{\de |\beta|}+3W\apic{A}=0.
\end{equation}
We can also take advantage of the Lorentz symmetry to show that the thermal coefficients
$\mathbb{S}$ and $\Gamma_w$ of the canonical spin tenor constitutive equation:
\begin{equation*}
\mean{ \frac{\I}{8}\bar{\Psi}\,\{\gamma^\lambda,\,[\gamma^\mu,\,\gamma^\nu]\,\}\,\Psi}
=\mathbb{S}\,\epsilon^{\lambda\mu\nu\rho}u_\rho+\Gamma_w\, \left( u^\lambda\varpi^{\mu\nu}+u^\nu\varpi^{\lambda\mu}+u^\mu\varpi^{\nu\lambda}\right)
+\mc{O}(\varpi^2),
\end{equation*}
satisfy the following relations
\begin{equation}
\label{eq:SpinTensAndSETRel}
\begin{split}
-\left(\frac{\mathbb{S}}{|\beta|}+\frac{\de \mathbb{S}}{\de|\beta|} \right)&= 2\mathbb{A}\apic{Can},\\
2\frac{\mathbb{S}}{|\beta|}&=\mathbb{W}_1\apic{Can}-\mathbb{W}_2\apic{Can},\\
\frac{\Gamma_w}{|\beta|}-\frac{\de\Gamma_w}{\de|\beta|} &=4\frac{\Gamma_w}{|\beta|} =G_1\apic{Can}-G_2\apic{Can},
\end{split}
\end{equation}
where $\mathbb{A}\apic{Can}$ and $\mathbb{W}_{1,2}\apic{Can}$ are the thermal coefficients of
Eq.~(\ref{setdecomp}) related to the mean value of \emph{canonical} stress-energy tensor.
Furthermore, because the axial current is dual to the spin tensor, we can show that
\begin{equation}
\label{eq:SpinTensAndAxialRel}
\mathbb{S}=\frac{1}{2}n\ped{A},\quad \Gamma_w=\frac{1}{2}W\apic{A}.
\end{equation}
Then, combining Eq.s (\ref{eq:SpinTensAndAxialRel}) and (\ref{eq:SpinTensAndSETRel}), the coefficients of canonical
stress-energy tensor and axial current are related by:
\begin{equation*}
\begin{split}
\mathbb{A}\apic{Can}=&-\left(\frac{n\ped{A}}{|\beta|}+\frac{\de n\ped{A}}{\de|\beta|} \right),\\
\frac{n\ped{A}}{|\beta|}=&\mathbb{W}_1\apic{Can}-\mathbb{W}_2\apic{Can},\\
\frac{W\apic{A}}{|\beta|}=&\frac{G_1\apic{Can}-G_2\apic{Can}}{2},
\end{split}
\end{equation*}
which expose an interesting connection between the Axial Vortical Effect (AVE) conductivity $W\apic{A}$
and the second order thermal coefficients of the canonical stress-energy tensor.

To understand the constraint~(\ref{AVERel}) and the relation between axial vortical effect and anomalies, we also consider
the case of a free massive field. In that case, axial current is not conserved, but its divergence is given by 
\begin{equation*}
\de_\mu\h{j}\ped{A}^\mu=2m\I \bar\Psi\gamma^5\Psi.
\end{equation*}
It follows that global equilibrium with a conserved axial charge can not be reached.
Then, to still use the previous global equilibrium analysis to massive field, we
simply set the axial chemical potential to zero and we consider global equilibrium
with just thermal vorticity and finite electric charge. As a consequence, the
symmetries impose that all chiral coefficients (i.e. those which are not parity
invariant) must be vanishing. However, the term in $W\apic{A}$ of axial current
decomposition is not chiral and consequently could be different from zero.
Since the conservation equation is changed, we expect that also the condition~(\ref{AVERel})
will be modified. We then have to consider the pseudo-scalar operator $\I\bar\Psi\gamma^5\Psi$
that appears on the divergence of axial current. Pseudoscalar thermal expectation
value can be decomposed at second order in thermal vorticity in the same way as other
local operators and we find that it is given by a single term:
\begin{equation*}
\mean{\I\bar\Psi\gamma^5\Psi}=(\alpha\cdot w)L^{\alpha\cdot w}
\end{equation*}
where the non chiral thermal coefficient can be obtained by
\begin{equation}
\label{eq:PseudoscalarCoeff}
L^{\alpha\cdot w}=\frac{1}{2}\mycorr{\{\h{K}_3,\h{J}_3\}\I\bar\Psi\gamma^5\Psi}.
\end{equation}
With this definition we find that the condition on axial vortical effect
conductivity $W\apic{A}$ becomes:
\begin{equation}
\label{AVERelMass}
|\beta|\frac{\de W\apic{A}}{\de |\beta|}+3W\apic{A}=-2mL^{\alpha\cdot w}.
\end{equation}
Differently from~(\ref{AVERel}) the constraint~(\ref{AVERelMass}) no longer imposes $W\apic{A}$
to be proportional to the third power of temperature and $W\apic{A}$ acquires terms which
depends on the mass of the fields.

As concluding remarks we give some results for these coefficients for the free massless Dirac field.
In that case this method reproduces the well-know~\cite{Kharzeev:2015znc} chiral vortical effect and
axial vortical effect conductivities
\begin{equation}
\label{eq:WVWA}
W\apic{V}= \frac{\mu\, \mu\,\ped{A} T}{\pi^2 },\quad
W\apic{A}= \frac{T^3}{6}+\frac{(\mu^2+\mu\ped{A}^2) T}{2 \pi^2}.
\end{equation}
In the case of massive Dirac fields,
global thermal equilibrium with thermal vorticity and vanishing axial chemical potential is well defined and the axial currents
mean value can be directed along the rotation of the fluid. In that situation the AVE conductivity for a free massive Dirac field
is~\cite{Buzzegoli:2017cqy}
\begin{equation}\label{coeffw2}
W^A=\frac{1}{2 \pi^2 |\beta| }\int_0^\infty \,\D p \left[n_F(E_p-\mu)+n_F(E_p+\mu)\right]\frac{2p^2+m^2}{E_p},
\end{equation}
where $E_p=\sqrt{p^2+m^2}$. This coefficient is related to pseudo-scalar thermal coefficient $L^{\alpha\cdot w}$ via Eq.~(\ref{AVERelMass})
and indeed pseudo-scalar coefficient is given by
\begin{equation}
\label{eq:PseudoL}
L^{\alpha\cdot w}=-\frac{m}{4 \pi^2 \beta^2 } \int_0^\infty \frac{\D p}{E_p} \left[n^{\prime}_F(E_p-\mu)+n^\prime_F(E_p+\mu)\right],
\end{equation}
where the prime on distribution functions stands for derivative respect to $E_p$.
We can give approximate results for integral in Eq.~(\ref{coeffw2}). For high temperature regime ($T\gg m$), if the
gas is non-degenerate ($|\mu|<m$), we extract the AVE conductivity behavior using the Mellin transformation technique~\cite{LandWeer}.
The result is
\begin{equation}
\label{eq:WAmassHighT}
\frac{W\apic{A}}{T}\simeq\frac{T^2}{6}+\frac{\mu^2}{2\pi^2}-\frac{m^2}{4\pi^2}-\frac{7\zeta'(-2)T^2}{8\pi^2}\left(\frac{m}{T}\right)^4+\mc{O}\left(\frac{m^6}{T^6}\right).
\end{equation}
The first term in mass was also obtained in~\cite{Flachi:2017vlp} where the axial
vortical effect was evaluated with the statistical operator~(\ref{eq:StatGEnoF}) but in curved space-time.
Low temperature behavior can also be extracted from~(\ref{coeffw2}),
see~\cite{Buzzegoli:2017cqy}. For a degenerate gas ($|\mu|>m$) at zero temperature we obtain\footnote{Notice that $W\apic{A}w^\mu\to(W\apic{A}/T) \vec{\omega}$,
	so there is no divergency for $T\to 0$.}:
\begin{equation*}
\frac{W\apic{A}}{T}=\frac{\mu^2}{2\pi^2}\frac{\sqrt{\mu^2-m^2}}{\mu}.
\end{equation*}
Instead for a non degenerate gas ($|\mu|<m$) at low temperature $T<<m$ we have
\begin{equation}
\label{eq:WAmassLowT}
W\apic{A}\approx\left(1+2\frac{T}{m}\right)\frac{(mT)^{3/2}}{\sqrt{2}\,\pi^{3/2}}\E^{|\beta|(\mu-m)}.
\end{equation}
Axial current corrections for rotating and accelerating fluids is also discussed in~\cite{Prokhorov:2017atp,Prokhorov:2018qhq}
for both massive and massless fields using an ansatz for Wigner function with thermal vorticity.

%*********************************************************************************
\section{Dirac Field in external electromagnetic field}
%*********************************************************************************
\label{sec:DiracInB}
Consider a Dirac field in external electromagnetic field.
The Lagrangian of the theory is given by
\begin{equation*}
\mathcal{L}= \frac{\I}{2}\left[ \bar{\Psi}\gamma^\mu \oraw{\de}_\mu \Psi -\bar{\Psi}\gamma^\mu \olaw{\de}_\mu \Psi\right]-m\bar{\Psi}\Psi-\h j^{\,\mu} A_\mu,
\end{equation*}
where $\h j^{\,\mu}=q\bar{\Psi}\gamma^\mu\Psi$, $q$ is the elementary electric charge of the field
and the gauge potential $A^\mu$ is an external non dynamic field.
This Lagrangian is obtained from the free Dirac one with the  minimal coupling 
substitution $\de_\mu\to\de_\mu+\I qA_\mu$  which ensures gauge invariance to the theory.
From Euler Eq.s we obtain the equations of motion (EOM) for the Dirac field:
\begin{equation*}
\slashed{\de}\Psi=-\I(q\slashed{A}+m)\Psi,\quad \slashed{\de}\bar{\Psi}=\bar{\Psi}\I(q\slashed{A}+m).
\end{equation*}
By applying Noether's theorem to this Lagrangian we obtain the operators
in Eq.~(\ref{eq:DiracOperator}).

%*********************************************************************************
\subsection{Symmetries in constant electromagnetic field}
%*********************************************************************************
\label{subsec:SymInFconst}
It is worth noticing that the symmetries of the theory of fermions in external electromagnetic field
are different from those of free fermions and from those of quantum electrodynamics. While a system
without external forces is symmetric for the full Poincaré group, some of the symmetries are lost
when external fields are introduced. Indeed, external fields do not transform together with the rest of the system.
In this section we discuss the symmetries of a system in the presence of an external constant homogeneous electromagnetic
field. We will examine the transformations that are still symmetries of the theory, the consequent
conserved quantities and the form of the generators of such transformations.

If the Lagrangian of our theory is invariant under translations, from Noether's theorem, we can identify
four operators. Those operators share three properties: they are conserved quantities, they are
the generators of translations and they constitute the four-momentum of the system. However, translation
invariance by itself does not guarantees that the same quantity must have all the three above properties
altogether. Consider again a system under an external electromagnetic field. In this situation, Poincaré symmetry
of space-time is broken. Only in the special case of a constant and homogeneous electromagnetic field
translation invariance is restored.
However, the Lagrangian is not invariant under space-time translation, but it acquires
a term that is a four-divergence. This term, under appropriate boundary conditions, does not affect the action
of the system and the overall invariance is preserved. Nevertheless, the consequence of the additional term is
that we can distinguish between three different operators, each of them having one of the three properties stated above.
This is understood with the Noether-Tassie-Buchdahl theorem~\cite{MR0180181,MR0180182,Bacry:1970ye}:
\textit{given a Lagrangian $\mathcal{L}(\Psi(x),\de_\mu\Psi(x),x)$ and the infinitesimal transformation
\begin{equation*}
x'^\mu=x^\mu+\delta x^\mu,\quad \Psi'=\Psi+\delta\Psi
\end{equation*}
such that  $\de_\mu\delta x^\mu=0$, which transforms the Lagrangian in
\begin{equation*}
\mathcal{L}(\Psi'(x'),\de'_\mu\Psi'(x'),x')=\mathcal{L}(\Psi(x),\de_\mu\Psi(x),x)+\de_\mu X^\mu,
\end{equation*}
where $X^\mu$ is a functional depending exclusively on $\Psi(x)$ and $x$, the quantity
\begin{equation*}
\Gamma^\mu=\frac{\delta \mathcal{L}}{\delta \de_\mu \Psi}\delta \Psi
-\left(\frac{\delta \mathcal{L}}{\delta \de_\mu \Psi}\de_\nu\Psi-\mathcal{L}\,g^\mu_{\,\nu} \right)\delta x^\nu-X^\mu
\end{equation*}
is conserved, i.e. divergence-less.}

Consider the Dirac Lagrangian in constant homogeneous electromagnetic field
\begin{equation*}
\mathcal{L}(\Psi(x),\de_\mu\Psi(x),x)=\bar\Psi(\I\slashed{\de}-m)\Psi-\h j^{\,\mu} A_\mu.
\end{equation*}
The translation transformation ($\delta\Psi=0$, $\delta x^\mu=\epsilon^\mu$) acts on the Dirac field but does not
act directly on the external gauge field. Therefore, a translation changes the Lagrangian by
\begin{equation*}
\begin{split}
\delta\mathcal{L}=&\mathcal{L}(\Psi'(x'),\de'_\mu\Psi'(x'),x')-\mathcal{L}(\Psi(x),\de_\mu\Psi(x),x)\\
=&\h j^{\,\mu} \de_\nu A_\mu \epsilon^\nu
=\h j^{\,\mu} (F_{\nu\mu}+\de_\mu A_\nu)\epsilon^\nu=-\h j^{\,\mu} (F_{\mu\nu}-\de_\mu A_\nu)\epsilon^\nu.
\end{split}
\end{equation*}
The quantity $X$ of Noether-Tassie-Buchdahl in this case is
\begin{equation*}
X^{\mu\nu}\equiv \h j^{\,\mu}( A^\nu + F^{\nu\lambda}x_\lambda).
\end{equation*}
Indeed its divergence is the variation of the Lagrangian
\begin{equation*}
\epsilon_\nu\de_\mu X^{\mu\nu}=(\de_\mu\h j^{\,\mu})( A^\nu + F^{\nu\lambda}x_\lambda)\epsilon_\nu
	+\h j^{\,\mu} (\de_\mu A_\nu + F_{\nu\mu})\epsilon^\nu=
-\h j^{\,\mu} (F_{\mu\nu}-\de_\mu A_\nu)\epsilon^\nu=\delta\mathcal{L}.
\end{equation*}
Therefore the theorem implies that the system has a canonical conserved tensor given by:
\begin{equation*}
\h\pi^{\,\mu\nu}\ped{can}=\h T^{\mu\nu}_0-\h j^{\,\mu} A^\nu-\h j^{\,\mu} F^{\nu\lambda} x_\lambda,
\end{equation*}
where $\h T^{\mu\nu}_0$ is the free canonical Dirac stress-energy tensor.
Using Belinfante procedure we can transform $\h T^{\mu\nu}_0-\h j^{\,\mu} A^\nu$ into the symmetric
stress-energy tensor of Dirac field in external magnetic field $\h T^{\mu\nu}_S$ and the above conserved
tensor can be written as:
\begin{equation}
\label{eq:GenTranInF}
\h\pi^{\,\mu\nu}\equiv\h T^{\mu\nu}_S-\h j^{\,\mu} F^{\nu\lambda} x_\lambda.
\end{equation}
From the above equation, we can simply verify that $\de_\mu \h\pi^{\,\mu\nu}=0$ form Eq.~(\ref{eq:OperRelF}).
Note that $\h{\pi}^{\,\mu\nu}$ is not symmetric and that it is gauge invariant.
The conserved quantities are obtained from the previous operators by
\begin{equation*}
\h\pi^{\,\mu}=\int \D^3 x\, \h{\pi}^{\,0\mu} 
\end{equation*}
and we can show that this four-vector constitutes the generators of the translation~\cite{Bacry:1970ye}.
However, the momentum of the system is still given by
\begin{equation*}
\h{P}^\mu=\int \D^3 x\, \h{T}^{0\mu} 
\end{equation*}
but it is no longer a conserved quantity and it is no longer the generator of translations.
Another difference with the four-momenta is that different components of this
vectors do not commute, instead they satisfy the commutation relation~\cite{Bacry:1970ye}
\begin{equation*}
[\h\pi^{\,\mu},\h\pi^{\,\nu}]=\I \h Q F^{\mu\nu},
\end{equation*}
where $\h{Q}$ is the electric charge operator.

As for Lorentz's transformations we expect that the variation of the Lagrangian is a full divergence
only for specific forms of transformations. For example, with a vanishing electric field and a constant
magnetic field, only the rotation along the direction of the magnetic field and the boost along the
magnetic field are symmetries of the theory. Therefore, only in these cases, the Lagrangian variation
could be vanishing or a full divergence.

Returning to the general case, by repeating the previous argument made for the translations
for the Lorentz transformation:
\begin{equation*}
\delta x^\mu=\omega^{\mu\nu}x_\nu,\quad \delta\Psi=-\frac{\I}{2}\omega_{\mu\nu}\sigma^{\mu\nu}\Psi,
\end{equation*}
we find that the transformed Lagrangian is
\begin{equation*}
\begin{split}
\delta\mathcal{L}=&
\omega^{\mu\nu}\h j^\lambda\l[\de_\lambda\l(x_\mu A_\nu-x_\nu A_\mu\r)+x_\mu F_{\nu\lambda}-x_\nu F_{\mu\lambda} \r].
\end{split}
\end{equation*}
We can show that the Lagrangian variation can also be written as
\begin{equation*}
\begin{split}
\delta\mathcal{L}=&\frac{1}{2}\omega^{\mu\nu}\de_\lambda\l[\h j^\lambda x_\mu \l(A_\nu-\frac{1}{2}x^\sigma F_{\sigma\nu}\r)
-\h j^\lambda x_\nu \l(A_\mu-\frac{1}{2}x^\sigma F_{\sigma\mu}\r)\r]\\
&-\frac{1}{2}x^\rho \h j^\lambda (\omega_{\lambda\sigma}F^\sigma_{\,\,\rho}-\omega_{\rho\sigma}F^\sigma_{\,\,\lambda}).
\end{split}
\end{equation*}
The first term of the r.h.s. is written as a four-divergence.  The remaining part can not be cast into
a four-divergence but  it is proportional to the following product
\begin{equation*}
(\omega\wedge F)_{\lambda\rho}=\omega_{\lambda\sigma}F^\sigma_{\,\,\rho}-\omega_{\rho\sigma}F^\sigma_{\,\,\lambda}.
\end{equation*}
The product of two non-vanishing anti-symmetric tensor of rank two, $\omega\wedge F$, is zero if and
only if $\omega$ is a linear combination of $F$ and its dual $F^*$ (or viceversa)~\cite{Bacry:1970ye}:
\begin{equation}
\label{eq:lemmaAntisymProd}
(\omega\wedge F)_{\lambda\rho}=0\quad\text{iff}\quad \omega_{\mu\nu}=k\,F_{\mu\nu}+k'\,F^*_{\mu\nu},\quad k,k'\in\mathbb{R}.
\end{equation}
Therefore, the part of Lagrangian variation which is not a divergence is vanishing when $\omega^{\mu\nu}$
is a linear combination of electromagnetic stress-energy tensor and its dual:
\begin{equation}
\label{eq:Lorentz_param_inv}
\omega^{\mu\nu}=a\,F^{\mu\nu}+\frac{b}{2}\epsilon^{\mu\nu\rho\sigma}F_{\rho\sigma}.
\end{equation}
This means, as expected, that the theory is invariant only under certain type of Lorentz transformations: the ones
generated with parameters of the form (\ref{eq:Lorentz_param_inv}).
For example, in the case of constant magnetic field, we recover
that the system is invariant only for rotation and boost along the magnetic field. Set then $\omega$ either as
$\omega_{\mu\nu}\propto F_{\mu\nu}$ or $\omega_{\mu\nu}\propto F^*_{\mu\nu}$, so that the Lagrangian variation is a four-divergence. In this case we can apply
Noether-Tassie-Buchdahl theorem and the two following quantities are divergence-less:
\begin{equation*}
\begin{split}
\h{\Gamma}^\lambda=&\frac{F^{\mu\nu}}{2}\l[ x_\mu \l(\h{T}^\lambda_{0\,\,\nu}-\h{j}^{\,\lambda} A_\nu+\frac{1}{2}\h{j}^{\,\lambda} x^\rho F_{\rho\nu}\r)\r.\\
	&\l.-x_\nu \l(\h{T}^\lambda_{0\,\,\mu}-\h{j}^{\,\lambda} A_\mu+\frac{1}{2}\h{j}^{\,\lambda} x^\rho F_{\rho\mu}\r)+\h{S}^\lambda_{\,\mu\nu}\r],\\
\h{\Gamma}^{*\lambda}=&\frac{F^{*\mu\nu}}{2}\l[ x_\mu \l(\h{T}^\lambda_{0\,\,\nu}-\h{j}^{\,\lambda} A_\nu+\frac{1}{2}\h{j}^{\,\lambda} x^\rho F_{\rho\nu}\r)\r.\\
	&\l.-x_\nu \l(\h{T}^\lambda_{0\,\,\mu}-\h{j}^{\,\lambda} A_\mu+\frac{1}{2}\h{j}^{\,\lambda} x^\rho F_{\rho\mu}\r)+\h{S}^\lambda_{\,\mu\nu}\r],
\end{split}
\end{equation*}
where $\h{S}^\lambda_{\mu\nu}$ is the canonical spin tensor of free Dirac field. After Belinfante transformation, the quantities become
\begin{equation}
\label{eq:LorentzInF}
\begin{split}
\h{\Gamma}^\lambda=&\frac{1}{2}F^{\mu\nu}\h{M}^\lambda_{\mu\nu},\quad
\h{\Gamma}^{*\lambda}=\frac{1}{2}F^{*\mu\nu}\h{M}^\lambda_{\mu\nu},\\
\h{M}^\lambda_{\mu\nu}\equiv&x_\mu \l(\h{T}^\lambda_{S\,\nu}+\frac{1}{2}\h{j}^{\,\lambda} x^\rho F_{\rho\nu}\r)
-x_\nu \l(\h{T}^\lambda_{S\,\mu}+\frac{1}{2}\h{j}^{\,\lambda} x^\rho F_{\rho\mu}\r)\\
=& x_\mu \l(\h{\pi}^{\,\lambda}_{\,\,\,\nu}-\frac{1}{2}\h{j}^{\,\lambda} x^\rho F_{\rho\nu}\r)
-x_\nu \l(\h{\pi}^{\,\lambda}_{\,\,\,\mu}-\frac{1}{2}\h{j}^{\,\lambda} x^\rho F_{\rho\mu}\r).
\end{split}
\end{equation}
We can define the integrals:
\begin{equation*}
\h{M}_{\mu\nu}=\int \D^3 x\,\h{M}^0_{\mu\nu},
\end{equation*}
which are conserved quantities only if contracted with $F^{\mu\nu}$ or $F^{*\mu\nu}$. The operators $\h{M}_{\mu\nu}$ are the generators of
Lorentz transformations if they are also a symmetry for the theory, otherwise the Wigner's theorem does not applies and
we can not say that such transformations admit an unitary and linear (or anti-unitary and anti-linear) representation.
For those operators, the following Algebra holds true~\cite{Bacry:1970ye}:
\begin{equation}\label{eq:MaxwellAlgebra}
\begin{split}
[\h{\pi}^{\,\mu},\h{\pi}^{\,\nu}]=&\I F^{\mu\nu}\h{Q},\\
\frac{1}{2}F_{\rho\sigma}[\h{\pi}{\,^\mu},\h{M}^{\rho\sigma}]=&\frac{\I}{2} F_{\rho\sigma}\l(\eta^{\mu\rho}\h\pi^{\,\sigma}-\eta^{\mu\sigma}\h\pi^{\,\rho}\r),\\
\frac{1}{2}F^*_{\rho\sigma}[\h{\pi}^{\,\mu},\h{M}^{\rho\sigma}]=&\frac{\I}{2} F^*_{\rho\sigma}\l(\eta^{\mu\rho}\h\pi^{\,\sigma}-\eta^{\mu\sigma}\h\pi^{\,\rho}\r),
\end{split}
\end{equation}
where $\h{Q}$ is the electric charge operator.
In the particular case of vanishing electric field and constant magnetic field along the $z$ axis, the Algebra becomes:
\begin{gather*}
[\h\pi_x,\,\h\pi_y]=\I|\vec{B}|\h{Q},\\
[\h{J}_z,\h\pi_x]=\I\h\pi_y,\quad [\h{J}_z,\h\pi_y]=-\I\h\pi_x,\\
[\h{K}_z,\h\pi_t]=-\I\h\pi_z,\quad [\h{K}_z,\h\pi_z]=-\I\h\pi_t.
\end{gather*}
%

%*********************************************************************************
\section{Chiral fermions in constant magnetic field}
%*********************************************************************************
\label{sec:ThermoInB}
Consider now a system consisting of free chiral fermions in an external homogeneous constant magnetic field $\vec{B}$
at global thermal equilibrium with vanishing thermal vorticity.
In this configuration, the chiral anomaly is vanishing because there is no electric field ($B\cdot E=0$).
It follows that global equilibrium without vorticity is described by a constant inverse four-temperature
$\beta$ and a constant axial chemical potential $\zeta\ped{A}$ (see Sec.~\ref{sec:GlobEqEM}).
Instead, the condition for the electric chemical potential reads
\begin{equation*}
\de^\mu \zeta(x)=F^{\nu\mu}\beta_\nu=\sqrt{\beta^2}F^{\nu\mu}u_\nu=\sqrt{\beta^2}E^\mu,
\end{equation*}
where $u$ is the fluid velocity directed along $\beta$ and $E$ is the comoving electric field.
Since we are considering the case without electric field
the global equilibrium condition is simply a constant $\zeta$.
The global equilibrium statistical operator then becomes
\begin{equation*}
\h{\rho} =  \frac{1}{Z} \exp \left[-\h P^\mu \beta_\mu +\zeta \h Q+\zeta\ped{A}\h Q\ped{A}  \right].
\end{equation*}
Notice that the operators $\h{P}^\mu$ are not the generators of translations, which are instead
given by $\h\pi^{\,\mu}$ and are obtained by integrating the conserved current
in Eq.~(\ref{eq:GenTranInF}). However, in the case of vanishing comoving electric field,
the projection of the inverse temperature along the four-momentum is equivalent to the projection
along of the generators of translations, that is:
\begin{equation*}
\h\pi^{\,\mu} \beta_\mu=\int_{\Sigma} \D \Sigma_\lambda \left(\h T^{\lambda\nu}-\h j^{\,\lambda} F^{\nu\sigma} x_\sigma\right)\beta_\nu
=\h P^\mu\beta_\mu-\sqrt{\beta^2}E^\sigma\int_{\Sigma} \D \Sigma_\lambda \h j^{\,\lambda} x_\sigma  =\h P^\mu\beta_\mu.
\end{equation*}
The statistical operator can now be written as
\begin{equation*}
\h{\rho} =  \frac{1}{Z} \exp \left[-\h \pi^{\,\mu} \beta_\mu +\zeta \h Q+\zeta\ped{A}\h Q\ped{A}  \right].
\end{equation*}
In this form, it is straightforward to use the algebra in Eq.~(\ref{eq:MaxwellAlgebra})
and translate the statistical operator of a quantity $a^\mu$. We find
\begin{equation*}
\h {\sf T}(a)\, \h\rho\, \h {\sf T}^{-1}(a) =\E^{\I a\cdot\h\pi}\h\rho\,\E^{-\I a\cdot\h\pi}
=\frac{1}{Z} \exp \left[-\h \pi^{\,\mu} \beta_\mu +\zeta \h Q+\zeta\ped{A}\h Q\ped{A}+a_\mu F^{\mu\nu}\beta_\nu\h{Q} \right].
\end{equation*}
Since $F^{\mu\nu}\beta_\nu$ is the comoving electric field, which is vanishing,
the statistical operator is homogeneous:
\begin{equation*}
\h {\sf T}(a)\, \h\rho\, \h {\sf T}^{-1}(a) =\h\rho .
\end{equation*}
%

%*********************************************************************************
\subsection{Exact thermal solutions}
%*********************************************************************************
\label{subsec:RitusMethod}
Having established the basic quantities of thermal equilibrium with a constant and homogeneous magnetic field,
we now move on to introduce the techniques of thermal field theory in order to find exact solutions for
thermodynamic equilibrium in a magnetic field. We start by giving a path integral description of the partition function.
Since the partition function is a Lorentz invariant, we can choose the evaluate it in the local rest frame where
 $u=(1,\vec{0})$.
In this frame, without loss in generality, the magnetic field is chosen along the $z$ axis and we adopt the Landau
gauge $A^\mu=(0,0,B x_1,0)$. The path integral formulation of the partition function in local rest frame
\begin{equation*}
\parz(T,\mu,\mu_5)=\tr\left[ \E^{-\beta (\h{H}-\mu \h{Q}-\mu\ped{A} \h{Q}\ped{A})}\right]
\end{equation*}
is given by\footnote{We added a mass term for generalization, although with mass we cannot have a conserved axial current.}
\begin{equation*}
\parz= C \int_{\Psi(\beta,\vec{x})=-\Psi(0,\vec{x})} \mathcal{D} \bar{\Psi}\,\mathcal{D} \Psi \,\, \exp \left( - S\ped{E}(\Psi,\bar{\Psi},\mu\ped{A}) \right)
\end{equation*}
where
the Euclidean action of Dirac fermions in external electromagnetic field is
\begin{equation*}
S\ped{E}(\Psi,\bar{\Psi},\mu_5)=\int_0^\beta\D\tau\int_{\vec{x}}
\bar{\Psi}(X)\left[\I(\gamma\cdot\pi^+)+m-\gamma_0 \gamma^5\mu\ped{A}  \right]\Psi(X)
\end{equation*}
and $\pi^+_\mu\equiv P^+_\mu-q A_\mu$, which is not to be confused with the generators of translations.

With regard to the exact solution, instead of solving the Dirac equation directly, we use the
Ritus method~\cite{Ritus:1972ky} (see~\cite{Leung:2005yq} for a brief recap of the method). The core concept
of Ritus method is that we can construct a complete set of orthonormal function, called $E_p$ Ritus functions,
such that the Euclidean action is rendered formally identical to the Euclidean action of a free Dirac field
in absence of external fields. The $E_p$ functions are constructed such that they are the matrix of the
contemporaneous eigenfunctions (eigenvectors) of the maximal set of mutually commuting operators
$\{(\gamma\cdot\pi)^2,\I\gamma_1\gamma_2,\gamma^5 \}$. From gamma algebra, it is straightforward to check that
\begin{equation*}
\I\gamma_1\gamma_2\delta(\sigma)=\sigma\Delta(\sigma),\quad
\frac{1+\chi\gamma^5}{2}\gamma^5=\chi\frac{1+\chi\gamma^5}{2},
\end{equation*}
with $\sigma=\pm$ and $\chi=\pm$ and we defined
\begin{equation*}
\Delta(\sigma)\equiv\frac{1+\I\sigma\gamma_1\gamma_2}{2}.
\end{equation*}
We can then show that~\cite{Ritus:1972ky,Leung:2005yq}
\begin{equation*}
(\gamma\cdot\pi^+)^2 E_{\h p \sigma}(X)={\underline{P}^+}^2 E_{\h p \sigma}(X)
\end{equation*}
where $\h p$ is a label for the quantum numbers $\{l,\omega_n,p_2,p_3\}$, the eigenvalues $\underline{P}^+$ are given by
\begin{equation*}
\underline{P}^+\equiv (\omega_n+\I\mu,0,-\bar\sigma\sqrt{2|qB|l},p_3),\quad \bar\sigma\equiv\text{sgn}(qB)
\end{equation*}
and the form of eigenfunction is
\begin{equation}
\label{eq:defEpsigma}
E_{\h p \sigma}(X)=N(n)\E^{\I(P_0 \tau + P_2 X_2+P_3 X_3)}D_n(\rho)
\end{equation}
where $N(n)=(4\pi |qB|)^{1/4}/\sqrt{n!}$ is a normalization factor, and $D_n(\rho)$ denotes the parabolic cylinder functions with argument
$\rho=\sqrt{2|qB|}(X_1 - p_2 /qB)$ and non-negative integer index $n = 0, 1, 2, \dots$ given by
\begin{equation*}
n=l+\frac{\sigma}{2}\text{sgn}(qB)-\frac{1}{2}.
\end{equation*}
Note that the form of the functions~(\ref{eq:defEpsigma}) strongly depends on the gauge chosen, in our case the Landau gauge.
Since the eigenfunction $E_{\h p \sigma}(X)$ does not depend on chirality, the maximal eigenfunctions of the operators
$\{(\gamma\cdot\pi)^2,\I\gamma_1\gamma_2,\gamma^5 \}$ are given by
\begin{equation}\label{eq:defE}
E_{\h p}(X)= \sideset{}{'}\sum_{\sigma=\pm} E_{\h p \sigma}(X)\Delta(\sigma),\quad
\bar{E}_{\h p}(X)=\gamma_0 E^\dagger_{\h p}(X)\gamma_0=\sideset{}{'}\sum_{\sigma'=\pm} E^*_{\h p \sigma'}(X)\Delta(\sigma')
\end{equation}
where the prime on the summation symbol denotes that the sum is subject to the constraint
\begin{equation*}
\sigma=\begin{cases} \text{sgn}(qB) & l=0\\ \pm & l>0 \end{cases}.
\end{equation*}
Some important properties can be derived from these definitions. Firstly, that the functions $E_p$ commute with $\gamma_0$ and with $\gamma^5$.
Secondly, they satisfy the orthogonality relation
\begin{equation*}
\int_X \bar{E}_{\h q}(X) E_{\h p}(X)=(2\pi)^4 \h\delta^{(4)}(\h p-\h q)\Pi(l)
\end{equation*}
where we defined
\begin{align*}
\delta^{(4)}(\h p-\h p')\equiv&\delta_{l,l'}\beta\delta_{\omega_n,\omega_{n'}}\delta(p_2-p_2')\delta(p_3-p_3')\\
\Pi(l)\equiv&\begin{cases} \frac{1+\I\bar\sigma\gamma_1\gamma_2}{2} & l=0\\ 1 & l>0 \end{cases}.
\end{align*}
And lastly, the action of the operator $(\gamma\cdot\pi^+)$ on these function is
\begin{equation*}
(\gamma\cdot\pi^+)E_{\h p}(X)=E_{\h p}(X)\gamma\cdot \underline{P}.
\end{equation*}

Since we showed that $E_p$ Ritus functions are a complete orthonormal functions, we can expand
the Dirac fields in these functions:
\begin{equation*}
\begin{split}
\Psi(X)=& T\sum_{\{\omega_n\}}\sum_{l=0}^\infty\int\D p_2\int \frac{\D p_3}{(2\pi)^3}E_{\h p}(X) \Psi(\underline{P})
\equiv \sumint_{\h P} E_{\h p}(X) \Psi(\underline{P}),\\
\bar{\Psi}(X)=&\sumint_{\h Q} \bar{\Psi}(\underline{Q}) \bar{E}_{\h q}(X).
\end{split}
\end{equation*}
Replacing this expansion on the Euclidean action, we find
\begin{equation*}
S\ped{E}(\Psi,\bar{\Psi},\mu_5)=\sumint_{\h P}\sumint_{\h Q}\int_{X}
\bar{\Psi}(\underline{Q})\bar{E}_{\h q}(X)\left[\I(\gamma\cdot\pi^+)+m-\gamma_0 \gamma^5\mu\ped{A}  \right]E_{\h p}(X)\Psi(\underline{P})
\end{equation*}
and, using the above mentioned properties of $E_p$ functions, we obtain
\begin{equation*}
S\ped{E}(\Psi,\bar{\Psi},\mu_5)=\sumint_{\h P}\bar{\Psi}(\underline{P})\Pi(l)\left[\I\gamma\cdot\underline{P}^+
+m-\gamma_0 \gamma^5\mu\ped{A}  \right]\Psi(\underline{P}).
\end{equation*}
Notice that this is formally identical to the Euclidean action of free Dirac field.

We can now proceed to evaluate the partition function. We first change the integration variables
in the partition function to the modes of $E_p$ functions $\bar{\Psi}(\underline{P})$ and $\Psi(\underline{P})$.
The partition function is then a Gaussian integral of Grassmann variables, whose result
is the exponent determinant. Hence the partition function becomes
\begin{equation}\label{eq_parzmagnetic}
\begin{split}
\parz=& \tilde{C} \int_{\Psi(\beta,\vec{x})=-\Psi(0,\vec{x})} \mathcal{D} \bar{\Psi}(\underline{P})\,\mathcal{D} \Psi(\underline{P})\times \\
&\times\exp \left\{ - \sumint_{\h P}\bar{\Psi}(\underline{P})\Pi(l)\left[\I\gamma\cdot\underline{P}^+ +m-\gamma_0 \gamma^5\mu\ped{A}  \right]\Psi(\underline{P}) \right\}\\
=&\tilde{C}\det\left[ \Pi(l)\left(\I\gamma\cdot\underline{P}^+ +m-\gamma_0 \gamma^5\mu\ped{A}  \right)\right].
\end{split}
\end{equation}
For the sake of clarity, for now on we will remove the factor $\Pi(l)$:
\begin{equation*}
\begin{split}
\parz=&\tilde{C}\det\left(\begin{array}{cc} m\idmat_{2\times 2}  & [\I(\omega_n+\I\mu)-\mu\ped{A}]\idmat_{2\times 2}+\sigma_i \underline{P}_i\\
(\I(\omega_n+\I\mu)+\mu\ped{A})\idmat_{2\times 2}-\sigma_i \underline{P}_i & m\idmat_{2\times 2}\end{array}\right).
\end{split}
\end{equation*}
The determinant is evaluated using the standard formula for block matrices
\begin{equation*}
\det\left(\begin{array}{cc} A & B\\ C & D\end{array}\right)= \det\left( AD-BD^{-1}CD\right);
\end{equation*}
replacing that into the partition function we have
\begin{equation*}
\begin{split}
\parz& =\tilde{C}\det\left[({\underline{P}^+}^2+m^2+\mu\ped{A}^2)\idmat_{2\times 2}-2\sigma_i \underline{P}_i\mu\ped{A} \right] \\
&= \tilde{C}\det\left[({\underline{P}^+}^2+m^2+\mu\ped{A}^2)^2-4|\vec{\underline{P}}|^2\mu\ped{A}^2 \right]\\
&= \prod_{\omega_n,l,p_3}\tilde{C}\left[({\underline{P}^+}^2+m^2+\mu\ped{A}^2)^2-4|\vec{\underline{P}}|^2\mu\ped{A}^2 \right],
\end{split}
\end{equation*}
where we evaluated the determinant as the product of the eigenvalues of the matrix.
To connect this quantity to the thermodynamics of the system, we are actually
interested in its logarithm:
\begin{equation}
\label{eq:logparzinB}
\log \parz=\sum_{\omega_n,l,p_3}\log\left[({\underline{P}^+}^2+m^2+\mu\ped{A}^2)^2-4|\vec{\underline{P}}|^2\mu\ped{A}^2 \right]+\text{cnst}.
\end{equation}
In the next subsection we evaluate the thermodynamic potential of the system from the
logarithm of partition function. Then, by simple derivation, we could obtain other
thermodynamic properties. However, starting from partition function in
Eq.~(\ref{eq_parzmagnetic}), we will obtain the thermal propagator of a chiral fermion
in a magnetic field. Once we have the propagator, we can use the point-splitting
procedure to evaluate other thermal properties that are not related to the
thermodynamic potential. We will use the thermal propagator to evaluate the mean value
of the electric current and of the axial current in the following subsections.

%--------------------------------------------------------------------------------
\subsection{Thermodynamic potential}
%--------------------------------------------------------------------------------
The thermodynamic potential $\Omega$ is derived from the partition function as following
\begin{equation*}
\Omega=\lim_{V\to\infty}-\frac{T}{V}\log \parz,
\end{equation*}
where the logarithm of partition function is given by Eq.~(\ref{eq:logparzinB}).
We can follow the usual techniques used for free fermions to evaluate the thermodynamic potential and to sum the Matsubara frequencies.
However, we must first consider that in this case the Landau levels generated by the magnetic field have different degeneracy factors
ad must be properly taken into account when performing the infinite volume limit. Let be $S$ the area in the $x-y$ plane and
$p_{\perp 1}$ and $p_{\perp 2}$ the momenta in that plane. In the limit of infinity area, the sum on modes becomes the following integrals:
\begin{equation*}
\lim_{S\to\infty}\frac{1}{S}\sum_{p_{\perp 1}}\sum_{p_{\perp 2}}=\int_{-\infty}^\infty\frac{\D p_{\perp 1}}{2\pi} \int_{-\infty}^\infty\frac{\D p_{\perp 2}}{2\pi}.
\end{equation*}
Each Landau level has a degeneracy associated with some quantum numbers; this degeneracy is gauge independent and it is given by:
\begin{equation*}
d_l=\left\lfloor \frac{|qB| S}{2\pi}\right\rfloor.
\end{equation*}
To obtain this degeneracy, we just have to evaluate the quantity
\begin{equation*}
\frac{\D p_{\perp 1}}{2\pi}\frac{\D p_{\perp 2}}{2\pi}
\end{equation*}
between two consecutive energy levels:
\begin{equation*}
d_l=\left\lfloor \frac{|qB| S}{2\pi}\right\rfloor=\int_l^{l+1}\frac{\D p_{\perp 1}}{2\pi}\frac{\D p_{\perp 2}}{2\pi}.
\end{equation*}
Therefore, removing the floor function, the infinite volume limit of the sum on the states of the system gives:
\begin{equation*}
\lim_{V\to\infty}\frac{1}{V}\sum_{\omega_n,l,p_3}=\frac{|qB|}{2\pi}\sum_{l=0}^\infty\int_{-\infty}^\infty\frac{\D p_3}{2\pi} \sum_{\{\omega_n\}}.
\end{equation*}
Consequently the thermodynamic potential reads:
\begin{equation*}
\begin{split}
\Omega=&\lim_{V\to\infty}-\frac{T}{V}\log \parz \\
=&-\frac{|qB|}{2\pi}\sum_{l=0}^\infty\int_{-\infty}^\infty\frac{\D p_3}{2\pi} 
T\sum_{\{\omega_n\}}\log\left[({\underline{P}^+}^2+m^2+\mu\ped{A}^2)^2-4|\vec{\underline{P}}|^2\mu\ped{A}^2 \right]+\text{cnst}.
\end{split}
\end{equation*}

The Matsubara sum can be performed as describe in~\cite{Laine:2016hma}.
The final result for thermodynamic potential is
\begin{equation*}
\Omega=-\frac{|qB|}{2\pi}\sum_{l=0}^\infty\sideset{}{'}\sum_{s=\pm}\int_{-\infty}^\infty\frac{\D p_3}{2\pi}
\left[E_s+T\sum_\pm\log\left(1+\E^{-\beta(E_s\pm\mu)}\right) \right]+\text{cnst},
\end{equation*}
where $E_s^2=[(p_3^2+2qBl)^{1/2}+s\mu\ped{A}]^2+m^2$ and the constraint of
$s=\bar{\sigma}$ for $l=0$ is caused by the projector $\Pi(l)$. This same
thermodynamic potential for chiral fermions in external magnetic field was used
in~\cite{Fukushima:2008xe} to derive the chiral magnetic effect (CME). This
expression can be used to obtain the electric and axial charge density, but instead
we are using the point-splitting procedure because the latter can also be used to
evaluate other thermodynamic functions related to currents. To do that, we first need
the thermal propagator of chiral fermions.

%--------------------------------------------------------------------------------
\subsection{Chiral fermion propagator in magnetic field}
%--------------------------------------------------------------------------------
Since we have used the Ritus method, the form of Euclidean action is formally
identical to those of a free Dirac field. It is therefore not surprising that the
fermionic propagator is obtained in the same way as the free case. The propagator in
Fourier modes can be obtained in path integral formulation by~\cite{Laine:2016hma}
\begin{equation*}
\mean{\tilde{\Psi}_a(P)\bar{\tilde\Psi}_b(Q)}_T=\frac{\int \Dpi \tilde\Psi \,\Dpi \tilde{\bar{\Psi}} \,\, \exp \l( - S\ped{E}\r) \tilde{\Psi}_a(P)\bar{\tilde\Psi}_b(Q)}{\int\Dpi \tilde\Psi \,\Dpi \tilde{\bar{\Psi}}\,\, \exp \l( - S\ped{E}\r)},
\end{equation*}
where in our case the form of the partition function is given in the first line of Eq.~(\ref{eq_parzmagnetic}):
\begin{equation*}
\begin{split}
\parz=& \tilde{C} \int_{\Psi(\beta,\vec{x})=-\Psi(0,\vec{x})} \mathcal{D} \bar{\Psi}(\underline{P})\,\mathcal{D} \Psi(\underline{P}) \,\,\times\\
&\times\exp \left\{ - \sumint_{\h P}\bar{\Psi}(\underline{P})\Pi(l)\left[\I\gamma\cdot\underline{P}^+ +m-\gamma_0 \gamma^5\mu\ped{A}  \right]\Psi(\underline{P}) \right\}.
\end{split}
\end{equation*}
The Grassmann integrals are straightforward and gives
\begin{equation*}
\mean{\bar{\Psi}(\underline{Q})_a \Psi(\underline{P})_b}=\delta^{(4)}(\underline{P}-\underline{Q})\mathcal{M}^{-1}_{ab}
\end{equation*}
where $a,b$ denotes spinorial indices and
\begin{equation*}
\mathcal{M}=\left[ \I\gamma\cdot \underline{P}^+ -\gamma_0\gamma^5\mu\ped{A}\right].
\end{equation*}
The inverse of $\mathcal{M}$ is easily written in terms of the projector into right and left chirality
states, which are defined by:
\begin{equation*}
\mathbb{P}_\chi=\frac{1+\chi\gamma_5}{2},\quad\text{i.e.}\quad
\mathbb{P}\ped{R}=\frac{1+\gamma_5}{2},\quad\mathbb{P}\ped{L}=\frac{1-\gamma_5}{2}.
\end{equation*}
We also introduce right and left chemical potential:
\begin{equation*}
\mu\ped{R}\equiv\mu+\mu\ped{A},\quad\mu\ped{L}\equiv\mu-\mu\ped{A},
\end{equation*}
and we define right and left charged momenta by
\begin{equation*}
P\ped{R/L}^\pm\equiv(\omega_n\pm\I \mu\ped{R/L},\vec{p}).
\end{equation*}
With this notation, after inverting $\mathcal{M}$, the thermal propagator is
\begin{equation*}
\mean{\bar{\Psi}(\underline{Q})_a \Psi(\underline{P})_b}=\delta^{(4)}(\underline{P}-\underline{Q})\sum_\chi
\left(\mathbb{P}_\chi\frac{-\I\slashed{\underline{P}}_\chi^+}{{\underline{P}^+_\chi}^2} \right)_{ab}.
\end{equation*}
This is the generalization in Euclidean space-time with chemical potentials of the propagator in \cite{Ritus:1972ky,Leung:2005yq}.
In the configuration space, the two-point function is
\begin{equation*}
\begin{split}
\mean{\bar{\Psi}(X)_a \Psi(Y)_b}&
=\sumint_{\h P} \sumint_{\h Q}\bar{E}_{\h p}(X)_{a'a} E_{\h q}(Y)_{bb'}\mean{\bar{\Psi}(\underline{P})_{a'} \Psi(\underline{Q})_{b'}}\\
&=-\sumint_{\h P} \sumint_{\h Q}\bar{E}_{\h p}(X)_{a'a} E_{\h q}(Y)_{bb'}\mean{\Psi(\underline{Q})_{b'} \bar{\Psi}(\underline{P})_{a'}}\\
&=-\sumint_{\h P} \sumint_{\h Q}\bar{E}_{\h p}(X)_{a'a} E_{\h q}(Y)_{bb'}\delta^{(4)}(\underline{P}-\underline{Q})\sum_\chi
\left(\mathbb{P}_\chi\frac{-\I\slashed{\underline{P}}_\chi^+}{{\underline{P}^+_\chi}^2} \right)_{b'a'}
\end{split}
\end{equation*}
where to go to second line we used fermion anti-commutation. Finally, integrating the delta we have
\begin{equation}
\label{eq:PropinB}
\mean{\bar{\Psi}(X)_a \Psi(Y)_b}=-\sumint_{\h P}\sum_\chi E_{\h p}(Y)_{bb'}
\left(\mathbb{P}_\chi\frac{-\I\slashed{\underline{P}}_\chi^+}{{\underline{P}^+_\chi}^2} \right)_{b'a'} \bar{E}_{\h p}(X)_{a'a}.
\end{equation}
%

%--------------------------------------------------------------------------------
\subsection{Electric current mean value}
%--------------------------------------------------------------------------------
\label{sec:CME}
Having derived the propagator, we now proceed to evaluate the mean value of electric current.
The following method is similar to the one used in~\cite{Fukushima:2009ft}, as we both use the Ritus method.
We take advantage of the point-splitting procedure to compute the thermal expectation value of electric current.
First, we write the current in Euclidean space-time and we split the coordinate point in which
the fields are evaluated as follows:
\begin{equation*}
\mean{\h{j}_\mu(X)}\!=\!(-\I)^{1-\delta_{\mu,0}}q\mean{\h{\bar{\Psi}}(X)\gamma_\mu \h{\Psi}(X)}
\!=\!\lim_{X_1,X_2\to X} (-\I)^{1-\delta_{\mu,0}}q\left(\gamma_\mu\right)_{ab}\mean{\h{\bar{\Psi}}_a(X_1)\h{\Psi}_b(X_2)}.
\end{equation*}
Then we plug the form of fermionic propagator~(\ref{eq:PropinB}) and we reconstruct the trace
on spinorial indices; eventually we obtain
\begin{equation*}
\mean{\h{j}_\mu(X)}=
-(-\I)^{1-\delta_{\mu,0}} q\sumint_{\h P}\sum_\chi \tr\left[\bar{E}_{\h p}(X)\, \gamma_\mu\, E_{\h p}(X)\,\mathbb{P}_\chi
\frac{-\I\slashed{\underline{P}}_\chi^+}{{\underline{P}^+_\chi}^2} \right].
\end{equation*}
It is convenient to indicate the components parallel to the magnetic field, which are the time component
and the $z$ component, with the parallel symbol ``$\parallel$''.
For those components the following commutator holds true:
\begin{equation*}
\left[ \gamma^\parallel_\mu\,, E_{\h p}(X)\right]=0.
\end{equation*}
Therefore, reminding the definitions of the Ritus $E_{\h p}(X)$ functions~(\ref{eq:defE}),
for the parallel component of electric current we obtain
\begin{equation*}
\begin{split}
\mean{\h{j}^{\,\parallel}_\mu(X)}
=&-(-\I)^{1-\delta_{\mu,0}} q\sum_{l=0}^\infty\int\frac{\D p_2}{2\pi}\int\frac{\D p_3}{(2\pi)^2}\sum_\chi T\sum_{\{\omega_n\}}
\sideset{}{'}\sum_{\sigma,\sigma'=\pm} E^*_{\h p\sigma'}(X)\, E_{\h p\sigma}(X)\times\\
&\times\tr\left[\Delta(\sigma')\,\Delta(\sigma)\,\gamma_\mu^\parallel\, \mathbb{P}_\chi\frac{-\I\slashed{\underline{P}}_\chi^+}{{\underline{P}^+_\chi}^2} \right].
\end{split}
\end{equation*}
We can simplify the previous expression by taking advantage of the following identity
\begin{equation*}
\Delta(\sigma)\Delta(\sigma')=\frac{1+\sigma\sigma'+\I(\sigma+\sigma')\gamma_1\gamma_2}{4}=\delta_{\sigma,\sigma'}\Delta(\sigma).
\end{equation*}
Notice that the dependence on $p_2$ is  only contained inside $E^*_{\h p\sigma'}(X) E_{\h p\sigma}(X)$.
We can then show that the integration on $p_2$ gives
\begin{equation*}
\int_{-\infty}^\infty \frac{\D p_2}{2\pi} E^*_{\h p\sigma'}(X) E_{\h p\sigma}(X)=|qB|\delta_{n,n'}.
\end{equation*}
Furthermore, it is convenient to split the sum on $l$ between the lowest Landau level (LLL) $l=0$
and the higher Landau levels (HLL) $l>1$. For $l=0$ the sums on $\sigma$ are
constrained to be equal to $\sigma=\sigma'=\bar\sigma=$sgn$(eB)$, and the momenta are given by $\underline{P}^+_\chi=(\omega_n+\I\mu_\chi,0,0,p_3)$; then at the lowest Landau level we have
\begin{equation*}
\mean{\h{j}^{\,\parallel}_\mu(X)}\ped{LLL}=-(-\I)^{1-\delta_{\mu,0}} q|qB|\int_{-\infty}^\infty\frac{\D p_3}{(2\pi)^2}\sum_\chi T\sum_{\{\omega_n\}}
\tr\left[\frac{1+\I\bar\sigma\gamma_1\gamma_2}{2}\gamma_\mu^\parallel\, \mathbb{P}_\chi\frac{-\I\slashed{\underline{P}}_\chi^+}{{\underline{P}^+_\chi}^2} \right].
\end{equation*}
After computing the trace, we find that the zero component is
\begin{equation*}
\begin{split}
\mean{\h{j}_0(X)}\ped{LLL}=&-\int_{-\infty}^\infty\frac{q|qB|\D p_3}{(2\pi)^2}\sum_\chi T\sum_{\{\omega_n\}}
\frac{-\I \left[(\omega_n+\I\mu_\chi)-\I p_3 \bar\sigma  \chi \right]}{(\omega_n+\I\mu_\chi)^2+p_3^2}\\
=&\int_{-\infty}^\infty\frac{q|qB|\D p_3}{(2\pi)^2}\sum_\chi T\sum_{\{\omega_n\}}
\frac{\I \left[(\omega_n+\I\mu_\chi)\right]}{(\omega_n+\I\mu_\chi)^2+p_3^2},
\end{split}
\end{equation*}
while the $z$ component is
\begin{equation*}
\begin{split}
\mean{\h{j}_3(X)}\ped{LLL}=&\I \int_{-\infty}^\infty\frac{q|qB|\D p_3}{(2\pi)^2}\sum_\chi T\sum_{\{\omega_n\}}
\frac{-\I \left[p_3+\I (\omega_n+\I\mu_\chi) \bar\sigma  \chi \right]}{(\omega_n+\I\mu_\chi)^2+p_3^2}\\
=& \int_{-\infty}^\infty\frac{\bar\sigma q|qB|\D p_3}{(2\pi)^2}\sum_\chi T\sum_{\{\omega_n\}}
\frac{\chi\I (\omega_n+\I\mu_\chi)}{(\omega_n+\I\mu_\chi)^2+p_3^2},
\end{split}
\end{equation*}
where the linear terms in $p_3$ were dropped because they are odd on $p_3$ and as such they vanish when integrated.
After the Matsubara sum we have
\begin{gather*}
\mean{\h{j}_0(X)}\ped{LLL}=q|qB|\sum_\chi\int_{-\infty}^\infty\frac{\D p_3}{(2\pi)^2}\frac{1}{2}\left[n\ped{F}(p_3-\mu_\chi)-n\ped{F}(p_3+\mu_\chi)\right],\\
\mean{\h{j}_3(X)}\ped{LLL}=q^2 B \sum_\chi\int_{-\infty}^\infty\frac{\D p_3}{(2\pi)^2}\frac{\chi}{2}\left[n\ped{F}(p_3-\mu_\chi)-n\ped{F}(p_3+\mu_\chi)\right].
\end{gather*}
Finally, taking advantage of the integral in $p_3$:
\begin{equation*}
\int_{-\infty}^\infty \frac{\D p_3 }{2}\left[n\ped{F}(p_3-\mu_\chi)-n\ped{F}(p_3+\mu_\chi)\right]=\mu_\chi
\end{equation*}
and summing on chiralities, we obtain
\begin{gather*}
\mean{\h{j}_0(X)}\ped{LLL}=\frac{q|qB|}{(2\pi)^2}(\mu\ped{R}+\mu\ped{L})=\frac{\mu q|qB|}{2\pi^2},\\
\mean{\h{j}_3(X)}\ped{LLL}=\frac{q^2 B}{(2\pi)^2}(\mu\ped{R}-\mu\ped{L})=\frac{q^2\mu\ped{A}}{2\pi^2}B.
\end{gather*}
Moving on now to the higher Landau levels, consider
\begin{equation*}
\begin{split}
\mean{\h{j}^{\,\parallel}_\mu(X)}\ped{HLL}=&-(-\I)^{1-\delta_{\mu,0}} q|qB|\sum_{l=1}^\infty\int_{-\infty}^\infty\frac{\D p_3}{(2\pi)^2}\sum_\chi T\sum_{\{\omega_n\}}
\sum_{\sigma,\sigma'=\pm}\delta_{n,n'}\times\\
&\times\tr\left[\Delta(\sigma')\,\Delta(\sigma)\,\gamma_\mu^\parallel\, \mathbb{P}_\chi\frac{-\I\slashed{\underline{P}}_\chi^+}{{\underline{P}^+_\chi}^2} \right].
\end{split}
\end{equation*}
When $l$ is fixed we can replace the $\delta_{n,n'}$ with the $\delta_{\sigma,\sigma'}$ and the
sum on $\sigma'$ becomes straightforward. The expression is similar
to the LLL case, we just have to replace $\bar\sigma$ with $\sigma$ and sum over $\sigma=\pm$.
Remind that now $\underline{P}$ has also a $y$ component.
After evaluating the trace and removing $p_3$ odd terms, we obtain
\begin{equation*}
\begin{split}
\mean{\h{j}_0(X)}\ped{HLL}
=&\sum_{l=1}^\infty\int_{-\infty}^\infty\frac{q|qB|\D p_3}{(2\pi)^2}\sum_\chi T\sum_{\{\omega_n\}}
\frac{2\I \left[(\omega_n+\I\mu_\chi)\right]}{{\underline{P}^+_\chi}^2},\\
\mean{\h{j}_3(X)}\ped{HLL}=&\sum_{l=1}^\infty\int_{-\infty}^\infty\frac{q|qB|\D p_3}{(2\pi)^2}\sum_\chi T\sum_{\{\omega_n\}}
\sum_{\sigma=\pm}\sigma\frac{\chi\I (\omega_n+\I\mu_\chi)+p_3}{{\underline{P}^+_\chi}^2}=0.
\end{split}
\end{equation*}
We found that the third component does not get corrections from HLL. Instead for the time component,
after the frequency sum, we obtain
\begin{equation*}
\begin{split}
\mean{\h{j}_0(X)}\ped{HLL}&=\sum_{l=1}^\infty\int_{-\infty}^\infty\frac{q|qB|\D p_3}{(2\pi)^2}\sum_\chi \left[n\ped{F}(E_{p_3,l}-\mu_\chi)-n\ped{F}(E_{p_3,l}+\mu_\chi)\right],
\end{split}
\end{equation*}
where $E_{p_3,l}\equiv\sqrt{p_3^2+2|qB|l}$.
For the perpendicular components ($\h{j}_x$ and $\h{j}_y$) we expect a vanishing result because they are
not allowed by the symmetries of the system. Indeed the explicit calculations confirmed this expectation.

In summary, restoring covariant expression, we found that the electric current has two thermodynamic function:
the electric charge density $n\ped{c}$ and the chiral magnetic effect (CME) conductivity $\sigma\ped{B}$:
\begin{equation*}
\mean{\h{j}_\mu(X)}=n\ped{c}\, u_\mu + \sigma\ped{B}\,B_\mu.
\end{equation*}
The electric charge density is given by the mean value $\mean{\h{j}_0(X)}$ at the local rest frame:
\begin{equation*}
\begin{split}
n\ped{c}=\frac{q|qB|}{2\pi^2}\Bigg\{\mu +\sum_{l=1}^\infty\int_{-\infty}^\infty\frac{\D p_3}{2}&
\Big[n\ped{F}(E_{p_3,l}-\mu\ped{R})-n\ped{F}(E_{p_3,l}+\mu\ped{R})+\\
&\hphantom{\Big[}+n\ped{F}(E_{p_3,l}-\mu\ped{L})-n\ped{F}(E_{p_3,l}+\mu\ped{L})\Big]\Bigg\},
\end{split}
\end{equation*}
while the CME conductivity is given by $\mean{\h{j}_3(X)}/B$, that is
\begin{equation}
\label{eq:CME_exact}
\sigma\ped{B}=\frac{q^2\mu\ped{A}}{2\pi^2}.
\end{equation}
To our knowledge the equation for the electric charge density of an electron gas in a magnetic medium
was first given in~\cite{Canuto:1969cs} and coincides with the expression above. The CME effect evaluated
here coincides with the one obtained with many other derivations~\cite{Kharzeev:2015znc}, however we want to
point out that this derivation is valid at thermal equilibrium, as the one in~\cite{Vilenkin:1980fu},
and that is non-perturbative in magnetic field.

%--------------------------------------------------------------------------------
\subsection{Axial current mean value}
%--------------------------------------------------------------------------------
We can compute the axial current mean value exactly as described above for the electric current.
Because of that we omit all the calculations.
The axial current constitutive equation is written in terms of an axial charge density $n\ped{A}$ and
a Chiral Separation Effect (CSE) conductivity $\sigma\ped{s}$:
\begin{equation*}
\mean{\h{j}_{A\mu}}=n\ped{A}\,u_\mu+\sigma\ped{s} B_\mu.
\end{equation*}
In this case too we found that only the lowest Landau level contributes to CSE
and that the final result is 
\begin{equation*}
\begin{split}
n\ped{A}=&\frac{|qB|}{2\pi^2}\Bigg\{\mu\ped{A}+\sum_{l=1}^\infty\int_{-\infty}^\infty\frac{\D p_3}{2} 
	\left[n\ped{F}(E_{p_3,l}-\mu\ped{R})-n\ped{F}(E_{p_3,l}+\mu\ped{R})+\right.\\
	&\left.-n\ped{F}(E_{p_3,l}-\mu\ped{L})+n\ped{F}(E_{p_3,l}+\mu\ped{L})\right]\Bigg\},\\
\sigma\ped{s}=&\frac{q\mu}{2\pi^2},
\end{split}
\end{equation*}
with $E_{p_3,l}=\sqrt{p_3^2+2|qB|l}$.
The last thermal coefficient is exactly the well known value of Chiral Separation Effect (CSE) conductivity~\cite{Kharzeev:2015znc}.

The same procedure can be followed to evaluate the axial charge density and the CSE conductivity of massive fermions with
vanishing axial chemical potential $\mu\ped{A}=0$. In that case, as discussed previously, the thermal equilibrium
can be reached and all the quantity discussed in this section are still well defined.
The results for the thermodynamic functions related to axial current are:
\begin{align}
n\ped{A}=&0, \nonumber \\
\label{eq:CSEMassive}
\sigma\ped{s}=&qB\int_{-\infty}^\infty \frac{\D p_3}{(2\pi)^2}\left[n\ped{F}(E_{p_3}-\mu)-n\ped{F}(E_{p_3}+\mu)\right],
\end{align}
with $E_{p_3}^2=p_3^2+m^2$. The CSE induces an axial current even if the system is not chiral.
It is apparent from the result above that CSE has an explicit mass dependence, as it is known that
it should have~\cite{Metlitski:2005pr}.
\newpage

%*********************************************************************************
\section{Constant vorticity and electromagnetic field}
%*********************************************************************************
\label{sec:ThermoInConstantB}
So far this contribution has focused on the general properties of the statistical operator
of global thermodynamic equilibrium with both vorticity and electromagnetic field~(\ref{eq:GEDO_EM}).
The following section will discuss the special case of a constant homogeneous electromagnetic field
($F^{\mu\nu}=$constant) for which we already studied the symmetries (Sec.~\ref{subsec:SymInFconst}).
As discussed in Sec.~\ref{sec:GlobEqEM}, global equilibrium can only be reached if condition~\eqref{eq_EqConstraintFmunu}
is satisfied, which in this case becomes
\begin{equation}
\label{eq:LieDerivFconst}
\mc{L}_\beta(F^{\mu\nu})=\varpi^\mu_{\,\,\sigma}F^{\sigma\nu}-\varpi^\nu_{\,\,\sigma}F^{\sigma\mu}\equiv(\varpi\wedge F)^{\mu\nu}=0.
\end{equation}
We already discussed this wedge product in Eq.~(\ref{eq:lemmaAntisymProd}). The Eq.~(\ref{eq:LieDerivFconst})
has two independent solutions: $F=k\varpi$ and $F=k'\varpi^*$, with $k$ and $k'$ real numbers.
In terms of the gauge potential the condition~(\ref{eq_EqConstraintAmu}) must be satisfied.
From Eq.~(\ref{eq:LieDerivFconst}), choosing the covariant gauge $A^\mu=\frac{1}{2}F^{\rho\mu}x_\rho$,
we find that condition~(\ref{eq_EqConstraintAmu})
is satisfied setting $\Phi=\frac{1}{2}b_\sigma F^{\sigma\lambda}x_\lambda$.
The equilibrium chemical potential~(\ref{eq_ZetaEmEq}) is then written as
\begin{equation}
\label{eq:zetaconstF}
\zeta(x)=\zeta_0-\beta_\sigma(x) F^{\lambda\sigma}x_\lambda+\frac{1}{2}\varpi_{\sigma\rho}x^\rho F^{\lambda\sigma}x_\lambda.
\end{equation}
The same solution can also be obtained by directly solving Eq.~(\ref{eq_ZetaGeneral}) using Eq.~(\ref{eq:LieDerivFconst}).
This last method to obtain the solution explicitly shows that the chemical potential in
Eq.~(\ref{eq:zetaconstF}) is not gauge dependent.
For constant magnetic field and vanishing thermal vorticity the solution (\ref{eq:zetaconstF}) reduces to $\zeta=$constant,
as it was correctly used in Sec.~\ref{sec:ThermoInB}.

Plugging the form (\ref{eq:zetaconstF}) inside the operator of Eq.~(\ref{eq:rhoLTEF}) we find:
\begin{equation*}
\h\rho=\frac{1}{\parz}\exp\l\{-\int\D\Sigma_\lambda \l[\l(\h{T}^{\lambda\nu}-\h{j}^{\,\lambda} F^{\nu\rho}x_\rho\r)\beta_\nu
-\frac{1}{2}\varpi_{\sigma\rho}\h{j}^{\,\lambda} x^\rho F^{\tau\sigma}x_\tau-\zeta_0\h{j}^{\,\lambda}\r] \r\}.
\end{equation*}
Inside the round bracket we recognize the divergence-less operator $\h\pi^{\lambda\nu}$ of Eq.~(\ref{eq:GenTranInF}),
whose integrals are the generators of translations. Expressing the coordinate dependence of $\beta$, we can then write
\begin{equation*}
\begin{split}
\h\rho=&\frac{1}{\parz}\exp\l\{-\int\D\Sigma_\lambda \l[\h{\pi}^{\lambda\nu}b_\nu
	+\varpi_{\nu\tau}x^\tau\h{\pi}^{\lambda\nu} -\frac{1}{2}\varpi_{\mu\nu}\h{j}^{\,\lambda} x^\nu F^{\tau\mu}x_\tau
	-\zeta_0\h{j}^{\,\lambda}\r] \r\}\\
=&\frac{1}{\parz}\exp\l\{-\int\D\Sigma_\lambda \l[\h{\pi}^{\lambda\nu}b_\nu
	+\varpi_{\mu\nu}x^\nu\l(\h{\pi}^{\lambda\mu}-\frac{1}{2}\h{j}^{\,\lambda} F^{\rho\mu}x_\rho\r)
	-\zeta_0\h{j}^{\,\lambda}\r] \r\}\\
=&\frac{1}{\parz}\exp\l\{-\int\D\Sigma_\lambda \l[\h{\pi}^{\lambda\nu}b_\nu
	-\frac{1}{2}\varpi_{\mu\nu}\l[x^\mu\l(\h{\pi}^{\lambda\nu}
	-\frac{1}{2}\h{j}^{\,\lambda} F^{\rho\nu}x_\rho\r)\r.\r.\r.\\
	&\l.\l.\l.-x^\nu\l(\h{\pi}^{\lambda\mu}-\frac{1}{2}\h{j}^{\,\lambda} F^{\rho\mu}x_\rho\r)\r]
	-\zeta_0\h{j}^{\,\lambda}\r] \r\};
\end{split}
\end{equation*}
this time we have recreated the divergence-less quantity $\varpi_{\mu\nu}\h{M}^{\lambda,\mu\nu}$ of Eq.~(\ref{eq:LorentzInF}) that generates the Lorentz transformations and that are symmetries of
the system. We can then integrate over the coordinate and we find:
\begin{equation*}
\h\rho=\frac{1}{\parz}\exp\left\{-b\cdot\h{\pi}+\frac{1}{2}\varpi:\h{M}+\zeta_0\h{Q}\right\}.
\end{equation*}
In the above form, the analogy with statistical operator without electromagnetic field in
Eq.~(\ref{eq:StatGEnoF}) is evident. In both cases the statistical operator is written with
the sum of conserved operators, each one weighted with a constant Lagrange multiplier.
Moreover, starting from a fixed point $x$ we can write the constants thermal fields as
\begin{equation*}
b_\mu=\beta(x)_\mu-\varpi_{\mu\nu}x^\nu,\quad
\zeta_0=\zeta(x)+\beta_\sigma(x) F^{\lambda\sigma}x_\lambda-\frac{1}{2}\varpi_{\sigma\rho}x^\rho F^{\lambda\sigma}x_\lambda,
\end{equation*}
from which the statistical operator becomes
\begin{equation*}
\begin{split}
\h\rho=&\frac{1}{\parz}\exp\Big\{-\beta(x)_\mu\left(\h{\pi}^\mu-F^{\lambda\mu}x_\lambda\h Q \right)+\\
&+\frac{1}{2}\varpi_{\mu\nu}\left(\h{M}^{\mu\nu}+x^\nu\h\pi^\mu-x^\mu\h\pi^\nu-x^\nu F^{\lambda\mu}x_\lambda\h{Q}\right)+\zeta(x)\h{Q}\Big\}.
\end{split}
\end{equation*}

It is important to point out that with an external magnetic field the Poincaré algebra is modified and becomes
the Algebra in Eq.~(\ref{eq:MaxwellAlgebra}), which we report here for convenience:
\begin{equation*}
\begin{split}
[\h{\pi}^\mu,\h{\pi}^\nu]=&\I F^{\mu\nu}\h{Q},\\
\frac{1}{2}F_{\rho\sigma}[\h{\pi}^\mu,\h{M}^{\rho\sigma}]=&\frac{\I}{2} F_{\rho\sigma}\l(\eta^{\mu\rho}\h\pi^\sigma-\eta^{\mu\sigma}\h\pi^\rho\r),\\
\frac{1}{2}F^*_{\rho\sigma}[\h{\pi}^\mu,\h{M}^{\rho\sigma}]=&\frac{\I}{2} F^*_{\rho\sigma}\l(\eta^{\mu\rho}\h\pi^\sigma-\eta^{\mu\sigma}\h\pi^\rho\r).
\end{split}
\end{equation*}
Notice that because $F$ is proportional to $\varpi$, if we replace $F$ with $\varpi$ and $F^*$ with $\varpi^*$
the last two algebra identities still hold true. Since the Algebra is known, we can translate
the statistical operator. Taking advantage of the unitary of the translation transformation,
the translated statistical operator is
\begin{equation*}
\begin{split}
\h {\sf T}(x)\, \h \rho\, \h {\sf T}^{-1}(x)
=&\frac{1}{\parz}\exp\Big\{-\h {\sf T}(x)\l(b\cdot\h{\pi}\r) \h {\sf T}^{-1}(x)+\\
&+\h {\sf T}(x)\l(\frac{\varpi:\h{M}}{2}\r) \h {\sf T}^{-1}(x)+\zeta_0\h {\sf T}(x)\,\h{Q}\, \h {\sf T}^{-1}(x)\Big\}.
\end{split}
\end{equation*}
Therefore we just need to evaluate how the operators $\h{\pi},\,\h{M} $ and $\h{Q}$
transform under translations. For a unitary transformation an operator $\h{K}$
transforms with
\begin{equation*}
\E^{\I\h{A}}\h K \E^{-\I\h{A}}\simeq \h{K}-\I\left[\h K,\h A\right]-\frac{1}{2}\left[\left[\h{K},\h{A}\right] ,\h{A}\right]
+\frac{\I}{6}\left[\left[\left[\h{K},\h{A}\right] ,\h{A}\right],\h{A}\right]+\cdots.
\end{equation*}
By applying this formula to our operators we obtain the complete transformation because after a certain order
all the commutators become vanishing. In particular, for the Lorentz transformation generators we find
\begin{equation*}
\frac{1}{2}\varpi_{\mu\nu}\h M^{\mu\nu}_x\equiv
\frac{1}{2}\varpi_{\mu\nu}\h {\sf T}(x) \h M^{\mu\nu} \h {\sf T}^{-1}(x)=\frac{1}{2}\varpi_{\mu\nu}\left(\h{M}^{\mu\nu}
+x^\nu\h\pi^\mu-x^\mu\h\pi^\nu-x^\nu F^{\lambda\mu}x_\lambda\h{Q}\right).
\end{equation*}
For the other operators instead we find:
\begin{equation*}
\h \pi^{\mu}_x\equiv \h {\sf T}(x)\, \h \pi^{\mu}\, \h {\sf T}^{-1}(x)=\h \pi^{\mu}-x_\rho F^{\rho\mu}\h{Q},\quad
\h {\sf T}(x)\, \h Q\, \h {\sf T}^{-1}(x)=\h Q.
\end{equation*}
With these definitions a translation transformation on the statistical operator
act as following:
\begin{equation*}
	\begin{split}
		\h {\sf T}(a)\,\h\rho\, \h {\sf T}^{-1}(a)=&\frac{1}{\parz}\exp\left\{-\beta(x)\cdot \h{\pi}_{x+a}+\frac{1}{2}\varpi:\h{M}_{x+a}+\zeta(x)\h{Q}\right\}\\
		=&\frac{1}{\parz}\exp\left\{-\beta(x-a)\cdot \h{\pi}_x+\frac{1}{2}\varpi:\h{M}_x+\zeta(x-a)\h{Q}\right\}.
	\end{split}
\end{equation*}
It follows that the statistical operator around a point $x$ can be written as:
\begin{equation}
\label{eq:GEDObetaF}
\h\rho=\frac{1}{\parz}\exp\left\{-\beta(x)\cdot \h{\pi}_x+\frac{1}{2}\varpi:\h{M}_x+\zeta(x)\h{Q}\right\}.
\end{equation}
%

%--------------------------------------------------------------------------------
\subsection{Expansion on thermal vorticity}
%--------------------------------------------------------------------------------
\label{sec:VortExpanF}
Following ref.~\cite{Buzzegoli:2018wpy} we use linear response theory to evaluate thermal expectation values
in the case of constant electromagnetic field. The purpose of this section is to give the thermal expectation
value of an operator $\h O$  at the point $x$ as a thermal vorticity expansion.
Using the properties of the trace, we can transfer the $x$ dependence from the operator $\h O$ to the
statistical operator~(\ref{eq:GEDObetaF}) written around the same point $x$:
\begin{equation*}
\begin{split}
\mean{\h{O}(x)}=&\frac{1}{\parz}\tr\l[\exp\left\{-\beta(x)\cdot \h{\pi}_x+\frac{1}{2}\varpi:\h{M}_x+\zeta(x)\h{Q}\right\}\h{O}(x)\r]\\
=&\frac{1}{\parz}\tr\l[\h {\sf T}(-x)\exp\left\{-\beta(x)\cdot \h{\pi}_x+\frac{1}{2}\varpi:\h{M}_x+\zeta(x)\h{Q}\right\}\h {\sf T}^{-1}(-x) \h{O}(0)\r]\\
=&\frac{1}{\parz}\tr\l[\exp\left\{-\beta(x)\cdot \h{\pi}+\frac{1}{2}\varpi:\h{M}+\zeta(x)\h{Q}\right\}\h{O}(0)\r].
\end{split}
\end{equation*}

To evaluate the mean value we expand the statistical operator of the last equality around vanishing vorticity.
First, we split the exponent of the statistical operator in two parts as follows:
\begin{equation*}
\h{\rho}=\dfrac{1}{\parz}\exp\left[\h{A}+\h{B}\right],\quad
\h{A}\equiv-\beta_\mu(x)\h{\pi}^\mu+\zeta(x)\h{Q},\quad\h{B}\equiv\frac{1}{2}\varpi:\h{M},
\end{equation*}
and then we expand on $\h{B}$, which is the part containing thermal vorticity. Since $\h{B}$ and $\h{A}$
satisfy the same algebra of the case discussed in~\cite{Buzzegoli:2018wpy}, the expansion will leads to the
same result, which is
\begin{equation}
\label{eq:VortExpEM}
\begin{split}
\mean {\h O(x)}=&\mean{\h O(0)}_{\beta(x)}-\alpha_\rho \mycorr{\,\h K^\rho \h O\,}-w_\rho \mycorr{\,\h J^\rho \h O\,}
+\frac{\alpha_\rho\alpha_\sigma}{2}\mycorr{\,\h K^\rho \h K^\sigma \h O\,}\\
&+\frac{w_\rho w_\sigma}{2} \mycorr{\,\h J^\rho \h J^\sigma \h O\,} +\frac{\alpha_\rho w_\sigma}{2}\mycorr{\,\{\h K^\rho,\h J^\sigma\}\h O\,}+\mathcal{O}(\varpi^3),
\end{split}
\end{equation}
where we defined
\begin{equation*}
\begin{split}
\mycorr{\h K^{\rho_1}\cdots \h K^{\rho_n} \h J^{\sigma_1}\cdots \h J^{\sigma_m}& \h O} \equiv
\int_0^{|\beta|} \frac{\D\tau_1\cdots\D\tau_{n+m}}{|\beta|^{n+m}}\times\\
&\times\mean{{\rm T}_\tau\left(\h K^{\rho_1}_{-\I \tau_1 u}\cdots\h K^{\rho_n}_{-\I \tau_n u}
	\h J^{\sigma_1}_{-\I \tau_{n+1} u} \cdots\h J^{\sigma_m}_{-\I \tau_{n+m} u} \h O(0)\right)}_{\beta(x),c}.
\end{split}
\end{equation*}
As discussed in Sec.~\ref{subsec:SymInFconst}, the boost and rotation defined starting from $\h{M}^{\mu\nu}$, i.e.
\begin{equation*}
	\h{K}^\mu=u_\lambda\h{M}^{\lambda\mu},\quad
	\h{J}^\mu=\frac{1}{2}\epsilon^{\alpha\beta\gamma\mu}u_\alpha\h{M}_{\beta\gamma},
\end{equation*}
are different from those of a system without external electromagnetic field. The other difference
with~\cite{Buzzegoli:2018wpy} is that in this case the averages $\mean{\cdots}_{\beta(x)}$
are made with the statistical operator
\begin{equation*}
\h\rho_0=\frac{1}{\parz_0}\exp\l\{-\beta(x)\cdot\h{\pi}+\zeta(x)\h{Q}\r\}.
\end{equation*}
%

%--------------------------------------------------------------------------------
\subsection{Currents and chiral anomaly}
%--------------------------------------------------------------------------------
\label{subsec:Feffect}
In this section we determine the constitutive equations for the electric and the axial current at first
order in thermal vorticity and we investigate the contributions from electric
and magnetic fields to the thermal coefficients related to vorticity. The constitutive
equations are obtained from the expansion on thermal vorticity given in the previous section.
Instead of a direct evaluation we use the conservation equations to
show that indeed no additional corrections from electric and magnetic field occur
to first order vorticous coefficients, such as the conductivity of the Chiral Vortical
Effect (CVE) and of the Axial Vortical Effect (AVE). In this way we obtain several
relations between those coefficients and their relation to the chiral anomaly.

Consider the case of global thermal equilibrium with constant vorticity $\varpi_{\mu\nu}$
and an electromagnetic field with strength tensor $F_{\mu\nu}=k\,\varpi_{\mu\nu}$,
with $k$ a constant.  It then follows that the comoving magnetic and electric fields
are parallel respectively to thermal rotation and thermal acceleration:
\begin{equation*}
B^\mu (x)=-k\,w^\mu(x),\quad E^\mu(x)=k\,\alpha^\mu(x).
\end{equation*}
For instance, in the case of a constant thermal vorticity caused by a rigid rotation 
along the $z$ axis and a constant magnetic field along $z$, we have:
\begin{equation*}
\varpi_{\mu\nu}=\frac{\Omega}{T_0}\l(\eta_{\mu 1}\eta_{\nu 2}-\eta_{\nu 1}\eta_{\mu 2}\r),\quad
F_{\mu\nu}=B\l(\eta_{\mu 1}\eta_{\nu 2}-\eta_{\nu 1}\eta_{\mu 2}\r),\quad
k=\frac{B T_0}{\Omega},
\end{equation*}
where $\Omega,\,T_0$ and $B$ are constants. In this example, electric and magnetic
fields are orthogonal and there is no chiral anomaly. However, in the general case,
the product $E\cdot B$ is non-vanishing. In that case, as we showed in Sec.~\ref{sec:GlobEqEM},
we can still discuss global equilibrium with chiral imbalance by defining a conserved Chern-Simons current.

By using the thermal vorticity expansion~(\ref{eq:VortExpEM})
we now proceed to write the thermal expectation
value of electric current at first order in thermal vorticity. We want to stress
that in the expansion~(\ref{eq:VortExpEM}) no approximations are made on the effects of the external electric
and magnetic fields; the expansion~(\ref{eq:VortExpEM}) only approximates the effects of vorticity.
At first order on thermal vorticity the only quantities that can contribute to the mean value of a current are the
four-vectors $w^\mu,\,\alpha^\mu$ and the scalars $E\cdot\alpha$,  $E\cdot w=-B\cdot\alpha$ and $B\cdot w$.
We therefore write the thermal expansion in terms of these quantities, which will define several thermal coefficients.
Taking into account the symmetries, the thermal vorticity expansion of the electric current is
\begin{equation}
\label{eq:ElecCurrEMVort}
\begin{split}
\mean{\,\h j^{\,\mu}(x)}=&\left[n\ped{c}^0+n\ped{c}^{E\cdot \alpha}(E\cdot \alpha)+n\ped{c}^{B\cdot w}(B\cdot w) \right]u^\mu
+W\apic{V} w^\mu+\sigma_E^{B\cdot \alpha} (B\cdot \alpha)E^\mu\\
&+\left[\sigma_B^0+\sigma_B^{E\cdot \alpha}(E\cdot \alpha)+\sigma_B^{B\cdot w}(B\cdot w) \right] B^\mu+\mc{O}\left(\varpi^2\right).
\end{split}
\end{equation}
Since the thermal coefficients $n\ped{c}^0$ and $\sigma_B^0$ must be evaluated at vanishing thermal
vorticity, they are exactly those computed in Sec.~\ref{sec:CME} (for vanishing electric field).
In particular, the Chiral Magnetic Effect (CME) conductivity at vanishing vorticity,
Eq.~(\ref{eq:CME_exact}), is
\begin{equation}
\label{eq:CME_repeat}
\sigma_B^0(x)=\frac{q\zeta_A}{2\pi^2|\beta(x)|}.
\end{equation}
All the other coefficients are related to thermal vorticity and they have the following
properties under parity, time-reversal and charge conjugation
\begin{equation}
\label{tab:ptcElectroMagnetic}
\begin{array}{lccc|cccc|ccc}
& E\cdot\alpha & E\cdot w & B\cdot w & n\ped{c}^0 & \sigma_B^0 & W\apic{V} & \sigma_E^0 & \sigma_B^{E\cdot\alpha} & \sigma_B^{B\cdot w} & \sigma_E^{B\cdot\alpha} \\ 
\hline
\group{P} & + & - & + & + & - & - & + & - & - & - \\
\group{T} & + & - & + & + & + & + & - & + & + & + \\
\group{C} & - & - & - & - & + & - & + & - & - & - \\
\end{array}
\end{equation}
Similarly the axial current thermal expectation value is
\begin{equation*}
	\begin{split}
		\mean{\,\h j^{\,\mu}\ped{A}(x)}=&\left[n\ped{A}^0+n\ped{A}^{E\cdot \alpha}(E\cdot \alpha)+n\ped{A}^{B\cdot w}(B\cdot w) \right]u^\mu
		+W\apic{A} w^\mu+\sigma_{sE}^{B\cdot \alpha} (B\cdot \alpha)E^\mu\\
		&+\left[\sigma_s^0+\sigma_s^{E\cdot \alpha}(E\cdot \alpha)+\sigma_s^{B\cdot w}(B\cdot w) \right] B^\mu+\mc{O}\left(\varpi^2\right).
	\end{split}
\end{equation*}
Each thermal coefficients is a function depending only on
\begin{equation}
\label{eq:LorScal}
|\beta|,\, \zeta,\, \zeta\ped{A},\, B^2,\, E^2,\, E\cdot B.
\end{equation}
The coordinate dependence of any thermal coefficients is completely contained inside
the Lorentz scalars in \eqref{eq:LorScal}.

With the constitutive equations written down, 
we are now looking for relations and constraints between those thermodynamic
coefficients. The conservation of electric current implies that
\begin{equation*}
\de_\mu\mean{\,\h j^{\,\mu}(x)}=\mean{\,\de_\mu\h j^{\,\mu}(x)}=0.
\end{equation*}
The coordinate derivative acts both on thermal coefficients and on thermodynamic fields.
We need to establish how the derivative acts on those quantities.
For thermodynamic fields, using the equilibrium conditions and the identities in
Appendix~\ref{sec:betaframeidentities}, we find
\begin{equation*}
\begin{split}
\de_\mu u^\mu=&0,\quad
\de_\mu w^\mu=-3\frac{w\cdot \alpha}{|\beta|},\quad
\de_\mu\alpha^\mu=\frac{2w^2-\alpha^2}{|\beta|},\\
\de_\mu B^\mu=&-3\frac{B\cdot \alpha}{|\beta|},\quad
\de_\mu E^\mu=-\frac{2(w\cdot B)+(\alpha\cdot E)}{|\beta|},\quad
\de_\mu(B\cdot \alpha)=0,\\
\de_\mu(E\cdot \alpha)=&-\frac{2}{|\beta|}\left[(w\cdot B)\alpha_\mu+(E\cdot w)w_\mu\right],\\
\de_\mu(B\cdot w)=&\frac{2}{|\beta|}\left[(w\cdot B)\alpha_\mu+(E\cdot w)w_\mu\right].
\end{split}
\end{equation*}
Moreover, we can also show that
\begin{equation*}
\begin{split}
\de_\mu |\beta|=-\alpha_\mu,\quad
\de_\mu\zeta=\beta^\nu F_{\nu\mu}=-|\beta| E_\mu,\quad
\de_\mu\zeta\ped{A}=0,\quad
\de_\mu(E\cdot B)=0,\\
\de_\mu|B|=-\frac{(E\cdot B)w_\mu+B^2\alpha_\mu}{|\beta||B|},\quad
\de_\mu|E|=-\frac{(E\cdot B)w_\mu+B^2\alpha_\mu}{|\beta||E|},
\end{split}
\end{equation*}
where 
\begin{equation*}
|\beta|=\sqrt{\beta^\sigma \beta_\sigma},\quad |B|=\sqrt{-B^\sigma B_\sigma},\quad
|E|=\sqrt{-E^\sigma E_\sigma}.
\end{equation*}
The derivative respect to coordinates of a thermodynamic function is
\begin{equation*}
\begin{split}
\de_\mu f(|\beta|,\zeta,\zeta_A,|B|,|E|,E\cdot B)=&\left(-\de_\mu|\beta| \frac{\de }{\de |\beta|}+\de_\mu\zeta \frac{\de }{\de \zeta}
+\de_\mu\zeta\ped{A} \frac{\de }{\de \zeta\ped{A}} +\de_\mu |B| \frac{\de }{\de |B|}\right.\\
&\left. +\de_\mu |E| \frac{\de }{\de |E|}+\de_\mu (E\cdot B) \frac{\de }{\de (E\cdot B)}\right) f.
\end{split}
\end{equation*}
Therefore, using the previous identities, the derivative of a thermodynamic function becomes
\begin{equation*}
\begin{split}
\de_\mu f=&\left[-\alpha_\mu\left( \frac{\de }{\de |\beta|}-\frac{|B|^2}{|\beta|}\frac{1}{|B|}\frac{\de }{\de |B|} -\frac{|B|^2}{|\beta|}\frac{1}{|E|}\frac{\de }{\de |E|} \right)\right.\\
&\left. -|\beta| E_\mu\frac{\de }{\de \zeta} -\frac{(E\cdot B)w_\mu}{|\beta|}\left( \frac{1}{|B|}\frac{\de }{\de |B|}+ \frac{1}{|E|} \frac{\de }{\de |E|}\right)\right] f.
\end{split}
\end{equation*}
We can also define the following short-hand notation:
\begin{equation*}
\de_{\tilde{\beta}}\equiv \frac{\de }{\de |\beta|}-\frac{|B|^2}{|\beta|}\de_{\tilde{B}},\quad
\de_{\tilde{B}}\equiv \frac{1}{|B|}\frac{\de }{\de |B|}+ \frac{1}{|E|} \frac{\de }{\de |E|},
\end{equation*}
from which the previous derivative is written as
\begin{equation*}
\de_\mu f=\left[-\alpha_\mu\de_{\tilde{\beta}}  -|\beta| E_\mu \de_\zeta -\frac{(E\cdot B)w_\mu}{|\beta|}\de_{\tilde{B}}\right] f.
\end{equation*}

We can now use the previous relations to impose electric current conservation by 
evaluating the divergence of the expansion in Eq.~(\ref{eq:ElecCurrEMVort}).
For the terms directed along the fluid velocity we find that no additional constraints
are required:
\begin{equation*}
\de_\mu\left(n^0 u^\mu\right)=\de_\mu\left(n^{E\cdot\alpha} (E\cdot \alpha) u^\mu\right)=\de_\mu\left(n^{B\cdot w} (B\cdot w) u^\mu\right)=0.
\end{equation*}
For the terms along the magnetic field we find:
\begin{equation*}
\begin{split}
\de_\mu\left(\sigma_B^0 B^\mu\right)=&-(B\cdot \alpha)\left[\frac{3}{|\beta|}+\de_{\tilde{\beta}} \right]\sigma_B^0
-(E\cdot B)|\beta|\de_\zeta \sigma_B^0\\&-\frac{(E\cdot B)(B\cdot w)}{|\beta|}\de_{\tilde{B}}\sigma_B^0,\\
\de_\mu\left(\sigma_B^{E\cdot\alpha}(E\cdot\alpha) B^\mu\right)=&-(B\cdot \alpha)(E\cdot\alpha)\left[\frac{3}{|\beta|}+\de_{\tilde{\beta}} \right]\sigma_B^{E\cdot\alpha}
-(E\cdot\alpha)(E\cdot B)|\beta|\de_\zeta \sigma_B^{E\cdot\alpha}\\
&-\frac{(E\cdot B)(B\cdot w)(E\cdot\alpha)}{|\beta|}\de_{\tilde{B}}\sigma_B^{E\cdot\alpha}\\
&-\frac{2}{|\beta|}\left[(w\cdot B)(B\cdot\alpha)+(E\cdot w)(B\cdot w) \right]\sigma_B^{E\cdot\alpha},\\
\de_\mu\left(\sigma_B^{B\cdot w}(B\cdot w) B^\mu\right)=& -(B\cdot \alpha)(B\cdot w)\left[\frac{3}{|\beta|}+\de_{\tilde{\beta}} \right]\sigma_B^{B\cdot w}
-(B\cdot w)(E\cdot B)|\beta|\de_\zeta \sigma_B^{B\cdot w}\\
&-\frac{(E\cdot B)(B\cdot w)^2}{|\beta|}\de_{\tilde{B}}\sigma_B^{B\cdot w}\\
&+\frac{2}{|\beta|}\left[(w\cdot B)(B\cdot\alpha)+(E\cdot w)(B\cdot w) \right]\sigma_B^{B\cdot w}.
\end{split}
\end{equation*}
Along electric field we have
\begin{equation*}
\begin{split}
\de_\mu\left(\sigma_E(B\cdot\alpha)E^\mu\right)=&-(B\cdot\alpha)(\alpha\cdot E)\left[\frac{1}{|\beta|}+\de_{\tilde{\beta}} \right]\sigma_E
-(B\cdot \alpha)E^2|\beta|\de_\zeta\sigma_E\\
&-\frac{(E\cdot B)(E\cdot w)}{|\beta|}\de_{\tilde{B}}\sigma_E.
\end{split}
\end{equation*}
Lastly, the divergence of the term along rotation is
\begin{equation*}
\de_\mu\left(W\apic{V} w^\mu\right)=-(w\cdot\alpha)\left[\frac{3}{|\beta|}+\de_{\tilde{\beta}} \right]W\apic{V}-(E\cdot w)|\beta|\de_\zeta W\apic{V}
-\frac{(E\cdot B)w^2}{|\beta|}\de_{\tilde{B}}W\apic{V}.
\end{equation*}

To impose that $\de_\mu \mean{\,\h j^{\,\mu}(x)}=0$ we sum all the previous pieces and
we split between the linear independent terms. Those terms must vanish independently of the
values of the electromagnetic field and of the thermal vorticity and several equalities are
obtained. Among those, we first consider the following identities:
\begin{equation*}
\begin{split}
\de_\zeta \sigma_B^0=&0,\quad
\de_{\tilde{B}} W\apic{V}=0,\quad
\de_\zeta \sigma_B^{E\cdot \alpha}=0,\quad
\de_{\tilde{B}} \sigma_B^{E\cdot \alpha}=0,\quad
\de_{\tilde{B}} \sigma_B^{B\cdot w}=0,\\
\de_\zeta \sigma_E^{B\cdot\alpha}=&0,\quad
\de_{\tilde{B}} \sigma_E^{B\cdot\alpha}=0.
\end{split}
\end{equation*}
Notice from the table in~(\ref{tab:ptcElectroMagnetic}) that $\sigma_B^{E\cdot \alpha}$
and $\sigma_E^{B\cdot\alpha}$ are related to $\group{C}$-odd correlator. Therefore they
must be odd functions of the electric chemical potential $\zeta$. But the previous
constraints require that they do not depend on $\zeta$, therefore they must be vanishing
\begin{equation*}
\sigma_B^{E\cdot \alpha}=0,\quad \sigma_E^{B\cdot\alpha}=0.
\end{equation*}
The previous constraints also require that $W\apic{V}$ and $\sigma_B^{B\cdot w}$
do not depend on $|B|$ and $|E|$\footnote{Note that to reduce the numbers of relations,
	we have indicated electric field and magnetic field derivative together with one derivative $\de_{\tilde{B}}$. However, electric and magnetic fields are independent and each derivative must be considered independently.}.
That is to say that the CVE conductivity $W\apic{V}$ is not affected by electric and magnetic
field. Moreover, electric current conservation also imposes that
\begin{equation*}
\begin{split}
\left(3+|\beta|\de_{\tilde{\beta}}\right)\sigma_B^0-|\beta|^2\de_\zeta W\apic{V}=&0,\\
\de_{\tilde{B}}\sigma_B^0+|\beta|^2\de_\zeta \sigma_B^{B\cdot w}=&0,\\
\left(3+|\beta|\de_{\tilde{\beta}}\right)\sigma_B^{B\cdot w}=&0,\\
\left(3+|\beta|\de_{\beta}\right)W\apic{V}=&0.
\end{split}
\end{equation*}
We can replace the known result for the CME conductivity $\sigma_B^0$ of Eq.~(\ref{eq:CME_repeat})
in the previous constraints to find find that
\begin{equation*}
\begin{split}
\de_\zeta W\apic{V}=&\frac{1}{|\beta|^2}\left(3+|\beta|\de_{\tilde{\beta}}\right)\sigma_B^0=\frac{q\zeta\ped{A}}{\pi^2|\beta|^3},\\
\de_\zeta \sigma_B^{B\cdot w}=&0,\\
\left(3+|\beta|\frac{\de}{\de\beta}\right)W\apic{V}=&0.
\end{split}
\end{equation*}
Again, since $\sigma_B^{B\cdot w}$ is $\group{C}$-odd it follows form second
equation that it must be vanishing
\begin{equation*}
\sigma_B^{B\cdot w}=0.
\end{equation*}

We want to empathize that we have found a relation between CVE and CME conductivities:
\begin{equation}
\label{eq:RelCME_CVE}
\de_\zeta W\apic{V}=\frac{1}{|\beta|^2}\left(3+|\beta|\de_{\tilde{\beta}}\right)\sigma_B^0.
\end{equation}
The CVE conductivity $W\apic{V}$ in Eq.~\eqref{eq:WVWA} satisfies this relation.
It is important to notice that Eq.~(\ref{eq:RelCME_CVE}) completely determines the CVE
conductivity from the CME one. Indeed, since $W\apic{V}$ is odd under charge
conjugation, by fixing the $\zeta$ part of $W\apic{V}$ we obtain the entire coefficient.
This also implies that the $\zeta$ part of $W\apic{V}$ is dictated by the chiral anomaly
as found in effective field theories~\cite{Sadofyev:2010is}.
Therefore, the CVE inherits all the properties proved for the CME. For instance, it
is known that the CME conductivity is completely dictated by the chiral anomaly~\cite{Kharzeev:2013ffa}
and that it is protected from corrections coming from interactions~\cite{Zakharov:2012vv,Feng:2018tpb}.
Since the relation~(\ref{eq:RelCME_CVE}) holds not only for a free theory but for any
microscopic interactions, as long as global thermal equilibrium is concerned, then
also the CVE conductivity is dictated by the chiral anomaly and it is universal.
Despite the CVE can be related to the vector-axial anomalous term of vector current
anomaly~\cite{Landsteiner:2012kd}, Eq.~(\ref{eq:RelCME_CVE}) shows that it can be
explained with just the electric charge conservation and the chiral anomaly.

Let us now move to the axial current. Similar steps can be followed to derive the
constraint equations between the thermal coefficients of axial current. In this case
we must impose the following identities between thermal expectation values:
\begin{equation*}
\de_\mu \mean{\,\h j^{\,\mu}\ped{A}(x)}= 2m\mean{\,\I\bar{\Psi}\gamma^5\Psi}-\frac{q^2(E\cdot B)}{2\pi^2},
\end{equation*}
where we also added the naive divergence term $ 2m\mean{\,\I\bar{\Psi}\gamma^5\Psi}$
which is due to the mass of the field. From symmetries the constitutive equation
for the pseudo-scalar is
\begin{equation*}
\begin{split}
\mean{\,\I\bar{\Psi}\gamma^5\Psi}=&L^{E\cdot B}(E\cdot B)+L^{\alpha\cdot w}(\alpha\cdot w)+L^{E\cdot w}(E\cdot w)\\
&+L^{(E\cdot B)\alpha^2}(E\cdot B)\alpha^2+L^{(E\cdot B)w^2}(E\cdot B)w^2\\
&+L^{(E\cdot w)(E\cdot \alpha)}(E\cdot w)(E\cdot\alpha)+L^{(E\cdot w)(B\cdot w)}(E\cdot w)(B\cdot w)+\mc{O}\left(\varpi^3\right).
\end{split}
\end{equation*}
The value of $L^{\alpha\cdot w}$ for free Dirac field has been reported
in~(\ref{eq:PseudoscalarCoeff}), and the other coefficients related to the
electromagnetic field can be computed with Ritus method.

Because the axial current is not conserved we find different identities compared to
the previous case of electric current. For instance, the identities related to the
Chiral Separation Effect (CSE) conductivity $\sigma_s^0$ and to the AVE conductivity
$W\apic{A}$ are
\begin{equation*}
\begin{split}
\left(3+|\beta|\frac{\de}{\de\beta}\right)W\apic{A}=&-2mL^{\alpha\cdot w},\\
\de_\zeta \sigma_s^0=&\frac{q^2}{2\pi^2|\beta|}-2mL^{E\cdot B},\\
\de_{\tilde{B}}\sigma^0_s=&-|\beta|^2\de_\zeta\sigma^{B\cdot w}_s,\\
\de_{\tilde{B}}W\apic{A}=&-2m|\beta|L^{(E\cdot B) w^2},\\
\de_\zeta W\apic{A}=&\frac{1}{|\beta|^2}\left(3+|\beta|\de_{\tilde{\beta}}\right)\sigma_s^0-\frac{2m}{|\beta|}L^{E\cdot w}.
\end{split}
\end{equation*}
The first equation has been discussed in~\cite{Buzzegoli:2018wpy} and in Sec.~\ref{subsec:NoF}.
The second equation is similar to the first: the first term  on the r.h.s. is coming
from the chiral anomaly and the second from the naive anomaly. Therefore for a
massive field, as discussed for the AVE, the CSE is not entirely dictated by the anomaly.
It is then not surprising to find a corrections to the CSE~\cite{Gorbar:2013upa} and
that it is affected by the mass, see Eq.~\eqref{eq:CSEMassive}. In the massive case
we also expect corrections from the external electromagnetic field both in the AVE
and in the CSE conductivities.

On the other hand, for massless field those constraints becomes
\begin{equation*}
\begin{split}
\de_\zeta \sigma_s^0=&\frac{q^2}{2\pi^2|\beta|},\\
\left(3+|\beta|\frac{\de}{\de\beta}\right)W\apic{A}=&0,\quad \de_{\tilde{B}}W\apic{A}=0, \\
\de_\zeta W\apic{A}=&\frac{1}{|\beta|^2}\left(3+|\beta|\de_{\beta}\right)\sigma_s^0.
\end{split}
\end{equation*}
In this case the CSE conductivity is completely fixed by the chiral anomaly as it is
clear by the first equation and by the fact that $\sigma_s^0$ must be an odd function
of $\zeta$. For the symmetries of axial current, $W\apic{A}$ has both terms which
depend on $\zeta$ and terms which depend only on $\zeta\ped{A}$ and $\beta$.
All these terms must satisfy the equations in the second line. In particular we
conclude that the AVE is not affected by the electromagnetic field. Moreover, from the
third line we see that the terms related to $\zeta$ are fixed by the CSE conductivity
and consequently they are dictated by the chiral anomaly. As it is evident from
previous discussion, this only occurs for the massless field.

In summary, by imposing the conservation equation we conclude that, at global
thermodynamic equilibrium with thermal vorticity and constant homogeneous
electromagnetic field, the chiral vortical effect is dictated by the chiral magnetic
effect. Then it is not affected by the mass of the particle, by the external
electromagnetic field or by radiative corrections. For the axial current this analysis
has showed that we need to distinguish between the massive and the massless case.
In the latter case we found that the chiral anomaly completely fixes the whole chiral
separation effect (CSE) but fixes only the part of axial vortical effect (AVE) conductivity
which depends on the electric chemical potential. We also found that the AVE is not
affected by the external electromagnetic field. For the massive case, despite it
exists a relation between the CSE and the AVE, both of them are affected by the mass
of the field, the external electromagnetic field and by radiative corrections.

\begin{acknowledgement}
I carried out part of this work while visiting Stony Brook University (New York, U.S.A.).
I would like to thank F. Becattini, E. Grossi and D. Kharzeev for stimulating discussions on the subject matter.
This research was supported in part by the Florence University with the fellowship
 ``Polarizzazione nei fluidi relativistici''.
\end{acknowledgement}

\appendix
%******************************************************************
\section*{Appendix: Thermodynamic relations in beta frame}
%******************************************************************
%\addcontentsline{toc}{section}{Appendix: Thermodynamic relations in beta frame}
\label{sec:betaframeidentities}
At global thermal equilibrium with thermal vorticity, thermodynamic fields satisfy several
equilibrium relations which constraints their coordinate dependence. In the $\beta$-frame
we can build several quantities from the four-vector $\beta$ and thermal vorticity~$\varpi$:
\begin{gather*}
u_\mu=\frac{\beta_\mu}{\sqrt{\beta^2}};\quad\Delta^{\mu\nu}=g^{\mu\nu}-u^\mu u^\nu;\quad
\varpi_{\mu\nu}=\de_\nu \beta_\mu = \epsilon_{\mu\nu\rho\sigma}w^\rho u^\sigma+\alpha_\mu u_\nu - \alpha_\nu u_\mu; \\
\alpha_\mu = \varpi_{\mu\nu} u^\nu; \quad w_\mu=-\frac{1}{2} \epsilon_{\mu\nu\rho\sigma}\varpi^{\nu\rho}u^\sigma; \quad
\gamma_\mu=(\alpha\cdot\varpi)^\lambda\Delta_{\lambda\mu}=\epsilon_{\mu\nu\rho\sigma} w^\nu \alpha^\rho u^\sigma.
\end{gather*}
Most of these quantities depend on coordinates and their derivatives are~\cite{Buzzegoli:2017cqy}:
\begin{gather*}
\de_\nu \beta_\mu = \varpi_{\mu\nu}; \quad
\de_\nu=-\alpha_\nu \frac{\de}{\de\sqrt{\beta^2}};\quad \varpi : \varpi=2\left(\alpha^2-w^2\right)\\
\de_\nu u_\mu=\frac{1}{\sqrt{\beta^2}}\big(\varpi_{\mu\nu}+\alpha_\nu u_\mu\big);\quad
\de^\alpha u_\alpha=0;\quad
u_\alpha\de^\alpha u_\mu=\frac{\alpha_\mu}{\sqrt{\beta^2}};\\
\de_\mu \alpha_\nu=\frac{1}{\sqrt{\beta^2}}\big( \varpi_{\nu\rho}\varpi^\rho_{\,\,\mu}+\alpha_\mu \alpha_\nu\big);\quad
\de^\alpha \alpha_\alpha=\frac{1}{\sqrt{\beta^2}}\big( 2w^2-\alpha^2\big);\quad
u_\alpha\de^\alpha \alpha^2=0;\\
\de_\mu w_\nu=\frac{1}{\sqrt{\beta^2}}\big(\alpha_\mu w_\nu -\frac{1}{2}\epsilon_{\nu\rho\sigma\lambda}\varpi^{\rho\sigma}\varpi^\lambda_{\,\,\mu}\big);\quad
\de^\alpha w_\alpha=-3\frac{w\cdot\alpha}{\sqrt{\beta^2}}; \quad u_\alpha\de^\alpha w^2=0;\\
\alpha^\sigma\de_\mu \alpha_\sigma=w^\sigma\de_\mu w_\sigma=\frac{1}{\sqrt{\beta^2}}\big(w^2\alpha_\mu-(\alpha\cdot w)w_\mu\big);\quad \de_\mu(\alpha\cdot w)=0;\\
\de_\alpha\gamma^\alpha=0;\quad \de^\alpha \Delta_{\alpha\beta}=-\frac{\alpha_\beta}{\sqrt{\beta^2}}.
\end{gather*}
%

%\bibliographystyle{spphys}
%\bibliography{MyBib}

\begin{thebibliography}{10}
	\providecommand{\url}[1]{{#1}}
	\providecommand{\urlprefix}{URL }
	\expandafter\ifx\csname urlstyle\endcsname\relax
	\providecommand{\doi}[1]{DOI \discretionary{}{}{}#1}\else
	\providecommand{\doi}{DOI \discretionary{}{}{}\begingroup
		\urlstyle{rm}\Url}\fi
	
	\bibitem{STAR:2017ckg}
	L.~Adamczyk, et~al., Nature \textbf{548}, 62 (2017).
	\newblock \doi{10.1038/nature23004}
	
	\bibitem{Kharzeev:2015znc}
	D.E. Kharzeev, J.~Liao, S.A. Voloshin, G.~Wang, Prog. Part. Nucl. Phys.
	\textbf{88}, 1 (2016).
	\newblock \doi{10.1016/j.ppnp.2016.01.001}
	
	\bibitem{Kharzeev:2007jp}
	D.E. Kharzeev, L.D. McLerran, H.J. Warringa, Nucl. Phys. A \textbf{803}, 227
	(2008).
	\newblock \doi{10.1016/j.nuclphysa.2008.02.298}
	
	\bibitem{Fukushima:2008xe}
	K.~Fukushima, D.E. Kharzeev, H.J. Warringa, Phys. Rev. \textbf{D78}, 074033
	(2008).
	\newblock \doi{10.1103/PhysRevD.78.074033}
	
	\bibitem{Heinz:2013th}
	U.~Heinz, R.~Snellings, Ann. Rev. Nucl. Part. Sci. \textbf{63}, 123 (2013).
	\newblock \doi{10.1146/annurev-nucl-102212-170540}
	
	\bibitem{HakimBook}
	R.~Hakim, \emph{Introduction to Relativistic Statistical Mechanics: Classical
		and Quantum} (World Scientific Publishing Co Inc, 2011)
	
	\bibitem{Israel:1978up}
	W.~Israel, Gen. Rel. Grav. \textbf{9}, 451 (1978).
	\newblock \doi{10.1007/BF00759845}
	
	\bibitem{Canuto:1969cs}
	V.~Canuto, H.Y. Chiu, Phys. Rev. \textbf{173}, 1220 (1968).
	\newblock \doi{10.1103/PhysRev.173.1220}
	
	\bibitem{Huang:2011dc}
	X.G. Huang, A.~Sedrakian, D.H. Rischke, Annals Phys. \textbf{326}, 3075 (2011).
	\newblock \doi{10.1016/j.aop.2011.08.001}
	
	\bibitem{Kovtun:2016lfw}
	P.~Kovtun, JHEP \textbf{07}, 028 (2016).
	\newblock \doi{10.1007/JHEP07(2016)028}
	
	\bibitem{Weickgenannt:2019dks}
	N.~Weickgenannt, X.L. Sheng, E.~Speranza, Q.~Wang, D.H. Rischke, Phys. Rev.
	\textbf{D100}, 056018 (2019).
	\newblock \doi{10.1103/PhysRevD.100.056018}
	
	\bibitem{Gao:2019znl}
	J.H. Gao, Z.T. Liang, Phys. Rev. \textbf{D100}, 056021 (2019).
	\newblock \doi{10.1103/PhysRevD.100.056021}
	
	\bibitem{Chen:2012ca}
	J.W. Chen, S.~Pu, Q.~Wang, X.N. Wang, Phys. Rev. Lett. \textbf{110}(26), 262301
	(2013).
	\newblock \doi{10.1103/PhysRevLett.110.262301}
	
	\bibitem{Zubarev1979}
	D.N. Zubarev, A.V. Prozorkevich, S.A. Smolyanskii, Theoretical and Mathematical
	Physics \textbf{40}(3), 821 (1979).
	\newblock \doi{10.1007/BF01032069}
	
	\bibitem{vanWeert1982}
	C.G. van Weert, Annals of Physics \textbf{140}, 133 (1982).
	\newblock \doi{10.1016/0003-4916(82)90338-4}
	
	\bibitem{BecaBetaF}
	F.~Becattini, L.~Bucciantini, E.~Grossi, L.~Tinti, Eur. Phys. J.
	\textbf{C75}(5), 191 (2015).
	\newblock \doi{10.1140/epjc/s10052-015-3384-y}
	
	\bibitem{Hayata:2015lga}
	T.~Hayata, Y.~Hidaka, T.~Noumi, M.~Hongo, Phys. Rev. \textbf{D92}(6), 065008
	(2015).
	\newblock \doi{10.1103/PhysRevD.92.065008}
	
	\bibitem{Hongo:2016mqm}
	M.~Hongo, Annals Phys. \textbf{383}, 1 (2017).
	\newblock \doi{10.1016/j.aop.2017.04.004}
	
	\bibitem{Becattini:2019dxo}
	F.~Becattini, M.~Buzzegoli, E.~Grossi, Particles \textbf{2}(2), 197 (2019).
	\newblock \doi{10.3390/particles2020014}
	
	\bibitem{Becattini:2018duy}
	F.~Becattini, W.~Florkowski, E.~Speranza, Phys. Lett. \textbf{B789}, 419
	(2019).
	\newblock \doi{10.1016/j.physletb.2018.12.016}
	
	\bibitem{BecCov}
	F.~Becattini, Phys. Rev. Lett. \textbf{108}, 244502 (2012).
	\newblock \doi{10.1103/PhysRevLett.108.244502}
	
	\bibitem{BecGro}
	F.~Becattini, E.~Grossi, Phys. Rev. \textbf{D92}, 045037 (2015).
	\newblock \doi{10.1103/PhysRevD.92.045037}
	
	\bibitem{Buzzegoli:2018wpy}
	M.~Buzzegoli, F.~Becattini, JHEP \textbf{12}, 002 (2018).
	\newblock \doi{10.1007/JHEP12(2018)002}
	
	\bibitem{Jensen:2013kka}
	K.~Jensen, R.~Loganayagam, A.~Yarom, JHEP \textbf{05}, 134 (2014).
	\newblock \doi{10.1007/JHEP05(2014)134}
	
	\bibitem{Ambrus:2014uqa}
	V.E. Ambruş, E.~Winstanley, Phys. Lett. \textbf{B734}, 296 (2014).
	\newblock \doi{10.1016/j.physletb.2014.05.031}
	
	\bibitem{Ambrus:2019cvr}
	V.E. Ambrus, E.~Winstanley, arXiv: 1908.10244  (2019)
	
	\bibitem{Becattini:2020qol}
	F.~Becattini, M.~Buzzegoli, A.~Palermo, arXiv:2007.08249  (2020)
	
	\bibitem{Becattini:2017ljh}
	F.~Becattini, Phys. Rev. \textbf{D97}(8), 085013 (2018).
	\newblock \doi{10.1103/PhysRevD.97.085013}
	
	\bibitem{Becattini:2019poj}
	F.~Becattini, D.~Rindori, Phys. Rev. \textbf{D99}(12), 125011 (2019).
	\newblock \doi{10.1103/PhysRevD.99.125011}
	
	\bibitem{Prokhorov:2019sss}
	G.Y. Prokhorov, O.V. Teryaev, V.I. Zakharov, Particles \textbf{3}(1), 1 (2020).
	\newblock \doi{10.3390/particles3010001}
	
	\bibitem{Prokhorov:2019cik}
	G.Y. Prokhorov, O.V. Teryaev, V.I. Zakharov, Phys. Rev. \textbf{D99}(7), 071901
	(2019).
	\newblock \doi{10.1103/PhysRevD.99.071901}
	
	\bibitem{Prokhorov:2019yft}
	G.Y. Prokhorov, O.V. Teryaev, V.I. Zakharov, JHEP \textbf{03}, 137 (2020).
	\newblock \doi{10.1007/JHEP03(2020)137}
	
	\bibitem{Buzzegoli:2017cqy}
	M.~Buzzegoli, E.~Grossi, F.~Becattini, JHEP \textbf{10}, 091 (2017).
	\newblock \doi{10.1007/JHEP07(2018)119, 10.1007/JHEP10(2017)091}.
	\newblock [Erratum: JHEP07,119(2018)]
	
	\bibitem{Hongo:2019rbd}
	M.~Hongo, Y.~Hidaka, Particles \textbf{2}(2), 261 (2019).
	\newblock \doi{10.3390/particles2020018}
	
	\bibitem{LandWeer}
	N.~Landsman, C.~van Weert, Physics Reports \textbf{145}(3), 141  (1987).
	\newblock \doi{http://dx.doi.org/10.1016/0370-1573(87)90121-9}
	
	\bibitem{Flachi:2017vlp}
	A.~Flachi, K.~Fukushima, Phys. Rev. \textbf{D98}(9), 096011 (2018).
	\newblock \doi{10.1103/PhysRevD.98.096011}
	
	\bibitem{Prokhorov:2017atp}
	G.~Prokhorov, O.~Teryaev, Phys. Rev. \textbf{D97}(7), 076013 (2018).
	\newblock \doi{10.1103/PhysRevD.97.076013}
	
	\bibitem{Prokhorov:2018qhq}
	G.~Prokhorov, O.~Teryaev, V.~Zakharov, Phys. Rev. \textbf{D98}(7), 071901
	(2018).
	\newblock \doi{10.1103/PhysRevD.98.071901}
	
	\bibitem{MR0180181}
	L.J. Tassie, H.A. Buchdahl, Austral. J. Phys. \textbf{17}, 431 (1964).
	\newblock \doi{10.1071/PH640431}
	
	\bibitem{MR0180182}
	H.A. Buchdahl, L.J. Tassie, Austral. J. Phys. \textbf{18}, 109 (1965).
	\newblock \doi{10.1071/PH650109}
	
	\bibitem{Bacry:1970ye}
	H.~Bacry, P.~Combe, J.L. Richard, Nuovo Cim. \textbf{A67}, 267 (1970).
	\newblock \doi{10.1007/BF02725178}
	
	\bibitem{Ritus:1972ky}
	V.I. Ritus, Annals Phys. \textbf{69}, 555 (1972).
	\newblock \doi{10.1016/0003-4916(72)90191-1}
	
	\bibitem{Leung:2005yq}
	C.N. Leung, S.Y. Wang, Nucl. Phys. \textbf{B747}, 266 (2006).
	\newblock \doi{10.1016/j.nuclphysb.2006.04.028}
	
	\bibitem{Laine:2016hma}
	M.~Laine, A.~Vuorinen, Lect. Notes Phys. \textbf{925}, pp.1 (2016).
	\newblock \doi{10.1007/978-3-319-31933-9}
	
	\bibitem{Fukushima:2009ft}
	K.~Fukushima, D.E. Kharzeev, H.J. Warringa, Nucl. Phys. \textbf{A836}, 311
	(2010).
	\newblock \doi{10.1016/j.nuclphysa.2010.02.003}
	
	\bibitem{Vilenkin:1980fu}
	A.~Vilenkin, Phys. Rev. \textbf{D22}, 3080 (1980).
	\newblock \doi{10.1103/PhysRevD.22.3080}
	
	\bibitem{Metlitski:2005pr}
	M.A. Metlitski, A.R. Zhitnitsky, Phys. Rev. \textbf{D72}, 045011 (2005).
	\newblock \doi{10.1103/PhysRevD.72.045011}
	
	\bibitem{Sadofyev:2010is}
	A.~Sadofyev, V.~Shevchenko, V.~Zakharov, Phys. Rev. D \textbf{83}, 105025
	(2011).
	\newblock \doi{10.1103/PhysRevD.83.105025}
	
	\bibitem{Kharzeev:2013ffa}
	D.E. Kharzeev, Prog. Part. Nucl. Phys. \textbf{75}, 133 (2014).
	\newblock \doi{10.1016/j.ppnp.2014.01.002}
	
	\bibitem{Zakharov:2012vv}
	V.I. Zakharov, Lect. Notes Phys. \textbf{871}, 295 (2013).
	\newblock \doi{10.1007/978-3-642-37305-3_11}
	
	\bibitem{Feng:2018tpb}
	B.~Feng, D.F. Hou, H.C. Ren, Phys. Rev. \textbf{D99}(3), 036010 (2019).
	\newblock \doi{10.1103/PhysRevD.99.036010}
	
	\bibitem{Landsteiner:2012kd}
	K.~Landsteiner, E.~Megias, F.~Pena-Benitez, Lect. Notes Phys. \textbf{871}, 433
	(2013).
	\newblock \doi{10.1007/978-3-642-37305-3_17}
	
	\bibitem{Gorbar:2013upa}
	E.V. Gorbar, V.A. Miransky, I.A. Shovkovy, X.~Wang, Phys. Rev. \textbf{D88}(2),
	025025 (2013).
	\newblock \doi{10.1103/PhysRevD.88.025025}
	
\end{thebibliography}

\end{document}